\documentclass[10pt,journal,a4paper]{IEEEtran}
\usepackage{latexsym}
\usepackage{textcomp}
\usepackage{amsfonts}
\usepackage{amsbsy}
\usepackage[tbtags]{amsmath}
\usepackage{amssymb}
\usepackage{times}
\usepackage{graphicx}
\usepackage{epstopdf}
\usepackage{caption2}
\usepackage{cite}
\usepackage{ctable}
\usepackage{longtable}
\usepackage{tabularx}
\usepackage{threeparttable}
\usepackage{verbatim}
\usepackage{multirow}
\usepackage{color}
\begin{document}
\newtheorem{lemma}{Lemma}
\newcommand{\notation}[1]{\emph{Notation:\ }}
\title{Robust Multi-Branch Tomlinson-Harashima Precoding in Cooperative MIMO Relay Systems}
\author{Lei Zhang,
Yunlong~Cai,
Rodrigo C. de Lamare,
   and     Minjian~Zhao

\thanks{Part of the manuscript was presented at the IEEE Vehicular Technology Conference-Spring, June 2013, Dresden, Germany.}
\thanks{L. Zhang, Y. Cai, and M. Zhao are with the Department of Information
Science and Electronic Engineering, Zhejiang University, Hangzhou 310027,
China (e-mail: bestleileisara@zju.edu.cn; ylcai@zju.edu.cn; mjzhao@zju.edu.cn).}
\thanks{R. C. de Lamare is with CETUC-PUC-Rio, 22453-900 Rio de Janeiro, Brazil,
and also with the Communications Research Group, Department of Electronics,
University of York, York Y010 5DD, U.K.
 (e-mail: rcdl500@ohm.
york.ac.uk).}
\thanks{This work was supported by the Fundamental Research Funds for the Central Universities
and the NSF of China under Grant 61101103.}
}

\maketitle

\begin{abstract}

This paper proposes the design of robust transceivers with
Tomlinson-Harashima precoding (THP) for multiple-input
multiple-output (MIMO) relay systems with amplify-and-forward (AF)
protocols based on a multi-branch (MB) strategy. The MB strategy
employs successive interference cancellation (SIC) on several
parallel branches which are equipped with different ordering
patterns so that each branch produces transmit signals by exploiting
a certain ordering pattern. For each parallel branch, the proposed
robust nonlinear transceiver design consists of THP at the source
along with a linear precoder at the relay and a linear
minimum-mean-squared-error (MMSE) receiver at the destination. By
taking the channel uncertainties into account, the source and relay
precoders are jointly optimised to minimise the mean-squared-error
(MSE). We then employ a diagonalization method along with some
attributes of matrix-monotone functions to convert the optimization
problem with matrix variables into an optimization problem with
scalar variables. We resort to an iterative method to obtain the
solution for the relay and the source precoders via
Karush-Kuhn-Tucker (KKT) conditions. An appropriate selection rule
is developed to choose the nonlinear transceiver corresponding to
the best branch for data transmission. Simulation results
demonstrate that the proposed MB-THP scheme is capable of
alleviating the effects of channel state information (CSI) errors
and improving the robustness of the system.

\end{abstract}


\IEEEpeerreviewmaketitle

\section{Introduction}

There has been considerable interest in wireless multiple-input
multiple-output (MIMO) communication systems, due to their potential
to enhance diversity and spectral efficiency \cite{MMIMO}. Recently,
MIMO techniques have been introduced in cooperative relay systems as
a means for further performance enhancement. It is well known that
relays are useful in increasing the coverage of wireless
communications under power and spectral constraints, and can provide
significant improvement in terms of both spectral efficiency and
link reliability. Amplify-and-forward (AF) is one of the most
popular relaying strategies due to low computational complexity and
small processing delay, where the relay simply processes the signals
received from the source without decoding and then forwards the
amplified signal to the destination. Therefore, using MIMO relays
with the AF strategy in multi-antenna relay systems has become a
very important
topic\cite{AFDF,relay,relayreceiver,joint,xing2010robust,jointrobust,erez2005capacity,costa1983writing,THPnodirect,THPdirect}.

Linear transceiver designs \cite{Lprec,switch,switchmc,Kek01,Kek02}
for dual-hop AF MIMO relay systems have been extensively
investigated in
\cite{relay,relayreceiver,joint,xing2010robust,jointrobust}. The
relay precoder in an AF-based MIMO relay system was first designed
in \cite{relay} to boost the overall channel capacity. In
\cite{relayreceiver}, a closed-form solution for the relay precoder
was proposed to minimize the mean-squared-error (MSE) in order to
improve the link quality. Joint design of the source and relay
precoding schemes was investigated in \cite{joint}, which can lead
to a better bit error ratio (BER) performance. Using the standard
Lagrange technique, the authors obtained the solution with an
iterative water-filling method. Also, both \cite{relayreceiver} and
\cite{joint} considered a linear minimum-mean-squared-error (MMSE)
destination equalizer. All the works above require the perfectly
known channel state information (CSI) in order to perform the
optimization. However, in practical relay systems, the CSI is
usually imperfect, since channel estimation errors are inevitable,
which should be taken into account in the transceiver design. In the
case that channel uncertainties are considered, some linear robust
techniques were proposed in \cite{xing2010robust} and
\cite{jointrobust}. The joint robust design of the linear relay
precoder and destination equalizer for a two-hop MIMO relay system
has been proposed  in \cite{xing2010robust}. More recently, by
taking source precoding into account the optimization of the source
and the relay processing matrices  using imperfect CSI was
investigated in \cite{jointrobust}.

As an alternative to the linear transceiver design, using nonlinear
precoding for MIMO relay channels has recently generated great
attention. A capacity achieving nonlinear dirty paper coding (DPC)
technique\cite{erez2005capacity} has been proposed for
presubtracting interference at the source prior to transmission.
Since DPC requires an infinite length of codewords and codebooks, it
is not suitable for practical use\cite{costa1983writing}. For this
reason, Tomlinson-Harashima precoding (THP) which originates from
DPC was presented as a low complexity alternative. This technique
employs modulo arithmetic and was originally proposed to combat
intersymbol interference (ISI) at the transmitter. In
\cite{THPnodirect}, Millar \emph{et al.} focused on the joint design
of linear processors for a two-hop network with THP employed at the
source. In \cite{THPdirect}, the direct link between the source and
the destination node was also considered. The authors proposed two
methods to solve the design problem, including a non-iterative
method to obtain the closed-form solutions for the precoders and an
iterative method to separately optimize the two precoders. Another
prominent precoding technique used in recent years is vector
perturbation (VP) viewed as a generalized THP
\cite{jimenez2012iterative,jimenez2011multiuser}, where the transmit
signal vector is perturbed by another vector to minimize transmit
power from the extended constellation. With the perturbation, a near
optimal performance is achieved by VP precoding. However, finding
the optimal perturbation vector can be a nondeterministic polynomial
time (NP)-hard problem. Conventional VP techniques based on sphere
encoding (SE) suffer from high computational complexity. Several
approaches have been reported in the last few years to reduce the
complexity of VP which include a tree-search method
\cite{barrenechea2012low} proposed as a low-complexity
implementation strategy. Considering the tradeoff between
computational complexity and performance, the THP algorithm is more
widely implemented in practical systems. For this reason, we focus
on THP in this paper.

To the best of our knowledge, few works have considered robust THP
in MIMO relay
systems\cite{millar2012robust,xing2012robust,tseng2012robust}.
In\cite{millar2012robust}, Millar \emph{et al.} employed some
approximations to relax the problem to make the optimization problem
tractable. A robust THP transceiver design for two-hop
non-regenerative MIMO relay networks with imperfect CSI was
presented. In \cite{xing2012robust}, a robust nonlinear design for a
multi-hop MIMO relay system was considered. In
\cite{tseng2012robust}, Tseng \emph{et al.} proposed THP with a
unitary precoder and adopted the primal decomposition to simplify
the optimization problem. Compared to prior work on robust
transceiver design that are only based on one particular
cancellation order and motivated by the sensitivity of THP to
channel uncertainty. In this paper, we propose a robust nonlinear
THP transceiver algorithm for MIMO relay systems in the presence of
imperfect CSI. Specifically, we consider a novel
successive-interference-cancellation (SIC) strategy for this system
based on a structure with multiple interference cancellation
branches. The original idea of this multi-branch (MB) strategy was
first proposed in
\cite{de2008multibranch,stmb,jidf,jio,mfsic,mfdf,mbdf,did} to
utilize the potential extra diversity gains for direct-sequence
code-division multiple access (DS-CDMA) systems and then extended to
precoding in \cite{MBTHP,multibranch2014}. This MB-SIC framework
consists of several SIC branches placed in parallel, and in each
branch a SIC scheme processes transmit signals with a given ordering
pattern\cite{MBIET}. For each branch, the nonlinear transceiver
design consists of a TH precoder at the source along with a linear
precoder used at the relay and a linear MMSE receiver at the
destination. We employ a diagonalization method along with some
attributes of matrix-monotone functions to obtain the optimal relay
and source precoders. The solution can be computed by using an
iterative method via the Karush-Kuhn-Tucker (KKT) conditions. An
appropriate selection rule is developed to choose the nonlinear
transceiver corresponding to the best branch for data transmission.
For every block prior to the data transmission, the source sends the
index of the selected optimal branch which is chosen by the
selection rule to the relay and the destination through a limited
feedforward channel. All the branches provide different capabilities
of interference cancellation for a given transmission block. Thus,
the best branch can be selected from them to obtain the best
possible performance. Simulation results demonstrate that the
proposed MB-THP scheme outperforms existing transceiver designs with
perfect and imperfect CSI. The contributions of this paper are
summarized as follows:

I) A novel robust MB-SIC strategy is developed according to different pre-stored ordering patterns for MIMO relay systems.

II) For each branch, we present the robust nonlinear transceiver design with THP using imperfect CSI.

III) We also propose a selection criterion for choosing the optimal branch corresponding to the minimum Euclidean distance for data transmission.

IV) Sub-optimal ordering schemes are developed to select a subset from the optimal ordering scheme set in a low-complexity way.

The rest of this paper is organized as follows. The proposed system model and channel error model are introduced in Section II. In Section III, we present the proposed robust MB-THP transceiver design for AF MIMO relay systems. The selection criterion, complexity analysis and the MB ordering schemes are described in Section IV. Simulation results and comparisons are given in Section V. Finally, conclusions are drawn in Section VI. Some technical details including the analysis are relegated to the Appendix.

\notation \ Throughout the paper, we denote vectors and matrices by lower and
upper case bold letters, respectively. ${\mathop{\rm E}\nolimits} \left[  \cdot  \right]$ stands for the statistical expectation.
The operators
${\left(  \cdot  \right)^T}$, ${\left(  \cdot  \right)^H}$, ${\left( \cdot \right)^{\rm{*}}}$, $\left|  \cdot  \right|$ and ${\mathop{\rm tr}\nolimits} \left(  \cdot  \right)$
denote the matrix transpose, Hermitian transpose, conjugate, determinant and trace, respectively.
The Kronecker product of matrices is denoted by $\otimes$.
  ${{\bf{A}}^{ - \frac{1}{2}}}$ represents the inverse square root of positive definite matrix ${\bf{A}}$.
  $\left\|  \cdot  \right\|$ is the Euclidean norm of the vector.
$\left\lfloor  \cdot  \right\rfloor$ represents the floor operator which returns the largest integer that is smaller than or
equal to the argument.

\section{System Model}
\subsection{Signal Model}
We consider a three-node AF MIMO relay communication system comprising of one source, one relay and one destination equipped with $N_s$, $N_r$ and $N_d$ antennas, respectively.
Due to long distance and possibly deep fading, the direct link between the source and destination is not considered in this paper.
In practice, this model is employed for the relay architectures of 3GPP LTE-Advanced\cite{peters2009relay}.
All the channels are assumed to be flat fading.

This system consists of a TH source precoder, a linear relay precoder and a linear MMSE receiver, as shown in Fig. 1. The quantity $\mathbf{s}$ is the ${N_d} \times 1$ input signal vector with zero mean and ${\rm{E}}\left[ {{\bf{s}}{{\bf{s}}^H}} \right] = \sigma _s^2{\bf{I}}$, where ${\bf{I}}$ denotes an identity matrix of appropriate dimension, and $\sigma _s^2$ is the average transmit power per antenna at the source. To ensure the transmission of ${N_d}$ independent data streams in this system, the number of transmit antennas should be larger than or equal to ${N_d}$, i.e., ${N_s} \ge {N_d}$. Each element of the transmit vector, ${\bf{s}} = [s_1 ,...,s_{N_d } ]^T$, is an $m$-ary square quadrature amplitude modulation (QAM) modulated signal, where the real and imaginary parts of $s_k$ belong to the set $\left\{ { \pm 1, \pm 3,..., \pm \left({\sqrt m  - 1} \right)} \right\}$. Then the input signal is sorted to generate multiple branch signals by the pre-designed cancellation ordering patterns. We introduce the ordering transformation matrix ${{\bf{T}}^{\left( l \right)}}$, $l \in \left\{ {1,...,L} \right\}$, which is a permutation matrix that has one entry of value equal to one, and corresponds to the ordering pattern employed in the $l$-th branch. The optimal ordering scheme conducts an exhaustive search with $L = N_s!$, where ! is the factorial operator. The reordered vector ${{\bf{\bar s}}^{(l)}}{\rm{ = }}{{\bf{T}}^{\left( l \right)}}{\bf{s}}$, which is based on the $l$-th cancellation order, is then recursively computed by a backward square matrix ${{\bf{C}}^{\left( l \right)}}$ for the $l$-th branch and a nonlinear modulo operation in order to perform a SIC operation.

\begin{figure*}[t]
\centering
\includegraphics[angle=270,scale=.55]{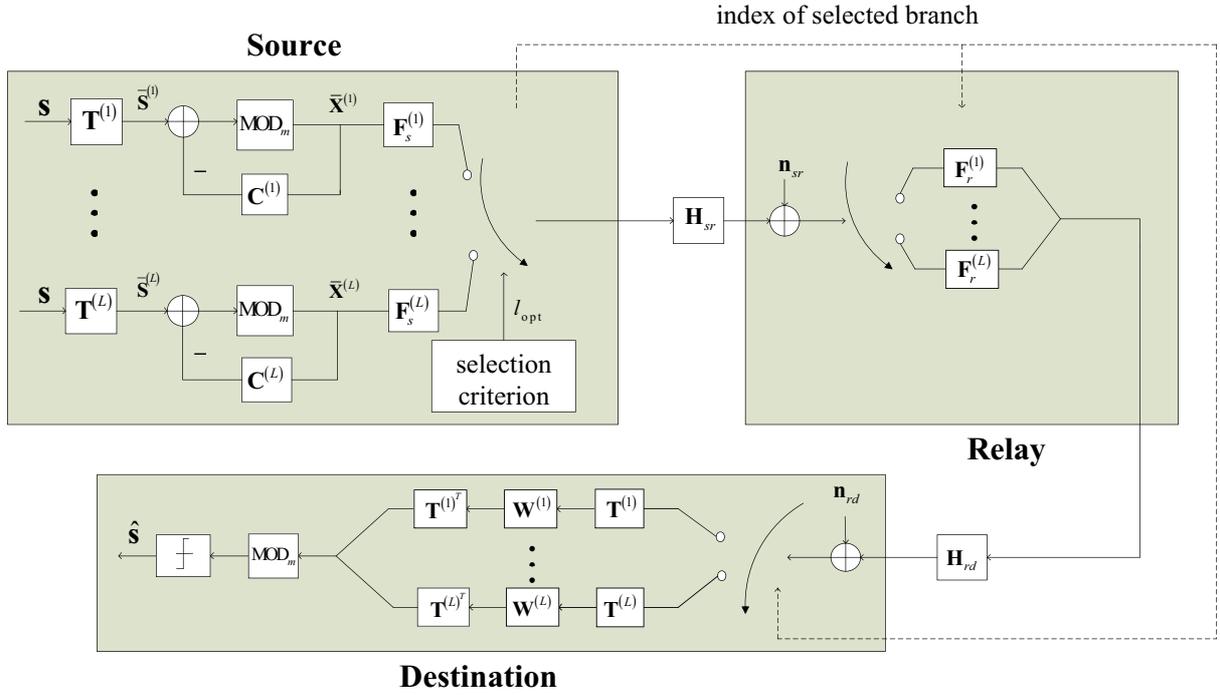}
\caption{{\small{MB-TH source and linear relay precoded AF MIMO relay system with MMSE receiver}}}
\end{figure*}

As shown in Fig. 1, ${\mathop{\rm MOD}\nolimits} _m \left( \cdot \right)$ stands for the modulo operator which is used to constrain a value to be within the region
$( - \sqrt m ,\sqrt m ]$. The modulo operator acts independently over the real and imaginary parts of its input according to the following rule
\begin{equation}
{\mathop{\rm MOD}\nolimits} _m \left( x \right) = x - 2\sqrt m \left\lfloor {\frac{{x + \sqrt m }}{{2\sqrt m }}} \right\rfloor.
\end{equation}

With ${{\bf{C}}^{\left( l \right)}}$ and the modulo operation in (1), the $l$-th branch channel symbols $\bar x_k^{(l)}$  are successively generated as
\begin{equation}
\bar x_k^{(l)} = \bar s_k^{(l)} - \sum\limits_{n = 1}^{k - 1} {{{\bf{C}}^{(l)}}\left( {k,n} \right)} \bar x_n^{(l)} + e_k^{(l)},
\end{equation}
\noindent where ${{\bf{C}}^{\left( l \right)}}$ is a strictly lower triangular matrix, ${{{\bf{C}}^{(l)}}\left( {k,m} \right)}$ is the element in the $k$th row and $n$th column of the matrix ${{\bf{C}}^{\left( l \right)}}$ and  ${{\bf{e}}^{\left( l \right)}} = {\left[ {e_1^{\left( l \right)},...,e_{{N_d}}^{\left( l \right)}} \right]^T}$ is the vector selected by the modulo operation to ensure that the real and imaginary parts of the elements in ${{{\bf{\bar x}}}^{\left( l \right)}}$ are bounded by the square region
 for the $l$-th branch. The Eq. (2) can be rewritten in matrix form as
\begin{equation}
{{{\bf{\bar x}}}^{\left( l \right)}} = {{\bf{U}}^{{{\left( l \right)}^{ - 1}}}}{{\bf{v}}^{\left( l \right)}},
\end{equation}
\noindent where  ${{\bf{v}}^{\left( l \right)}} = {{{\bf{\bar s}}}^{\left( l \right)}} + {{\bf{e}}^{\left( l \right)}}$,
${{\bf{U}}^{\left( l \right)}}\in {\mathop{\rm \mathbb{C}}\nolimits} ^{N_d  \times N_d }$ is a lower triangular matrix with ones on the main diagonal and it is given by
${{\bf{U}}^{\left( l \right)}} = {{\bf{C}}^{\left( l \right)}} + {\bf{I}}$ . As a result of the modulo operation, the elements of ${{{\bf{\bar x}}}^{\left( l \right)}}$ are no longer uncorrelated and uniformly distributed over the Voronoi region\cite{fischer2002precoding}. This leads to the $l$-th branch channel symbol vector ${{{\bf{\bar x}}}^{\left( l \right)}}$ having slightly higher energy than ${{{\bf{\bar s}}}^{\left( l \right)}}$. This slight
increase in the average energy is termed precoding loss\cite{fischer2002precoding}. For moderate to high $m$ this energy increase can be neglected\cite{THPnodirect,windpassinger2004precoding,simeone2004linear}, thus we still have ${\rm{E}}\left[ {{{{\bf{\bar x}}}^{\left( l \right)}}{{{\bf{\bar x}}}^{{{\left( l \right)}^H}}}} \right] = \sigma _s^2{\bf{I}}$.  Based on a selection criterion, the optimum source precoder, relay precoder and receiver corresponding to the ${l_{\rm{opt}}}$-th branch are chosen for data transmission. Then the source sends the index of the selected optimal branch to the relay and the destination through a limited feedforward channel before data transmission. The signal transmission is carried out in two stages. In the first phase, the signal is processed by the selected precoding matrix ${\bf{F}}_{s}^{\left( {l_{\rm{opt}}} \right)}  \in {\mathop{\rm\mathbb{C}}\nolimits} ^{N_s  \times N_d }$ for the ${l_{\rm{opt}}}$-th branch. The received signal ${\bf{y}}_r^{\left( {l_{\rm{opt}}} \right)}$ corresponding to the ${l_{\rm{opt}}}$-th cancellation order at the relay is given by
\begin{equation}
{\bf{y}}_r^{\left( {l_{\rm{opt}}} \right)} = {{\bf{H}}_{sr}}{\bf{F}}_s^{\left( {l_{\rm{opt}}} \right)}{{{\bf{\bar x}}}^{\left( {l_{\rm{opt}}} \right)}} + {{\bf{n}}_{sr}},
\end{equation}
\noindent where ${\bf{H}}_{sr}  \in {\mathop{\rm \mathbb{C}}\nolimits} ^{N_r  \times N_s }$ denotes the MIMO channel matrix between the source and the relay. The vector ${\bf{n}}_{sr}$ is the additive noise component at the relay which is modeled as a circularly symmetric complex Gaussian random vectors with zero-mean and correlation matrix ${\rm{E}}\left[ {{{\bf{n}}_{sr}}{\bf{n}}_{sr}^H} \right] = \sigma _{{n_{sr}}}^2{\bf{I}}$, where $\sigma _{n_{sr} }^2$  is the average noise power at the relay.

In the second phase, the relay forwards the received signals to the destination after performing linear precoding which corresponds to the selected branch, while the source does not transmit data. Thus, the ordered signal ${\bf{y}}_d^{\left( {l_{\rm{opt}}} \right)}$ received at the destination corresponding to the ${l_{\rm{opt}}}$-th branch is given by
\begin{align}
&{\bf{y}}_d^{\left( {l_{\rm{opt}}} \right)} = {{\bf{T}}^{\left({l_{\rm{opt}}} \right)}}{{\bf{H}}_{rd}}{\bf{F}}_r^{\left( {l_{\rm{opt}}} \right)}{{\bf{H}}_{sr}}{\bf{F}}_s^{\left( {l_{\rm{opt}}}\right)}{{{\bf{\bar x}}}^{\left( {l_{\rm{opt}}} \right)}}\nonumber\\
& \quad \quad \quad \quad  + {{\bf{T}}^{\left( {l_{\rm{opt}}} \right)}}{{\bf{H}}_{rd}}{\bf{F}}_r^{\left( {l_{\rm{opt}}} \right)}{{\bf{n}}_{sr}}{\rm{ + }}{{\bf{T}}^{\left( {l_{\rm{opt}}}\right)}}{{\bf{n}}_{rd}},
\end{align}
\noindent where ${{\bf{T}}^{\left( {{l_{{\rm{opt}}}}} \right)}}$
represents the selected ordering transformation matrix, ${\bf{F}}_{r}^{\left( {l_{\rm{opt}}}\right)}  \in {\mathop{\rm\mathbb{C}}\nolimits} ^{N_r  \times N_r }$ is the selected relay precoder for the ${l_{\rm{opt}}}$-th branch. ${\bf{H}}_{rd}  \in {\mathop{\rm \mathbb{C}}\nolimits} ^{N_d  \times N_r }$ stands for the MIMO channel matrix between the relay and the destination. Mathematically, the equivalent channel matrix after a specific transmit pattern can be denoted as ${\bf{H}}_{rd}^{\left( {l_{\rm{opt}}} \right)} = {{\bf{T}}^{\left( {l_{\rm{opt}}} \right)}}{{\bf{H}}_{rd}}$. By transforming the channel matrix, the columns of the channel matrix ${{\bf{H}}_{rd}}$ are permutated\cite{liu2007improved}. The vector ${\bf{n}}_{rd}$ is the zero-mean complex Gaussian noise vector at the destination with ${\rm{E}}\left[ {{{\bf{n}}_{rd}}{\bf{n}}_{rd}^H} \right] = \sigma _{{n_{rd}}}^2{\bf{I}}$,  where $\sigma _{n_{rd} }^2$  denotes the received average noise power at the destination.

At the destination, the selected linear receiver ${{\bf{W}}^{\left( {l_{\rm{opt}}} \right)}}$ is then employed to detect the received signal. The detected signal is given by:
\begin{equation}
{{{\bf{\hat v}}}^{\left({l_{\rm{opt}}} \right)}} = {{\bf{W}}^{\left( {l_{\rm{opt}}} \right)}}{\bf{y}}_d^{\left( {l_{\rm{opt}}} \right)}.
\end{equation}

Then the signal can be re-arranged in the original order by using ${{\bf{T}}^{\left( {l_{\rm{opt}}} \right)}}$. Thus, the final output is obtained by
 \begin{equation}
{\bf{\hat s}}{\rm{ = Q}}\left( {{\rm{MOD}}\left( {{{\bf{T}}^{{{\left( {l_{\rm{opt}}} \right)}^T}}}{{{\bf{\hat v}}}^{\left( {l_{\rm{opt}}}\right)}}} \right)} \right),
\end{equation}
\noindent where ${\mathop{\rm Q}\nolimits} \left(  \cdot  \right)$ denotes the quantization operation.

\subsection{Channel Error Model}

Since perfect CSI is typically not available at the transmitter and at the receiver \cite{tseng2012robust}, robust methods have been developed to deal with imperfect CSI. To model the channel matrix distribution, the well-known Kronecker model is adopted for the covariance of the CSI mismatch \cite{xing2010robust}. Although we focus on the
channel error model based on the CSI mismatch, we note that it can be easily extended to the model with channel feedback, since the work in \cite{dabbagh2008multiple} has built a relationship between them and verified that they are equivalent problems. We have the following expressions:
\begin{align}
&{\bf{H}}_{sr}  = {\bf{\bar H}}_{sr}  + \Delta {\bf{H}}_{sr},\\
&{\bf{H}}_{rd}  = {\bf{\bar H}}_{rd}  + \Delta {\bf{H}}_{rd},
\end{align}
\noindent where ${\bf{\bar H}}_{sr}$ and ${\bf{\bar H}}_{rd}$ are the estimated channel matrices, while $\Delta {\bf{H}}_{sr}$ and $\Delta {\bf{H}}_{rd}$ are the corresponding channel estimation error matrices. $\Delta {\bf{H}}_{sr}$ can be written as {$\Delta {\bf{H}}_{sr}  = {\bf{\Sigma }}_{sr}^{1/2} {\bf{H}}_{i.i.d} {\bf{\Psi }}_{sr}^{T/2}$}, and
$\Delta {\bf{H}}_{rd}$ can be written as $\Delta {\bf{H}}_{rd}  = {\bf{\Sigma }}_{rd}^{1/2} {\bf{H}}_{i.i.d} {\bf{\Psi }}_{rd}^{T/2}$,
 where the elements of ${\bf{H}}_{i.i.d}$ are independent and identically distributed Gaussian random variables with zero mean and unit variance. Both the relay and destination have the estimated CSI. Thus, $\Delta {\bf{H}}_{sr}$  and $\Delta {\bf{H}}_{rd}$ have the matrix-variate complex Gaussian distribution, which can be expressed as \cite{matrixdistribution}
\begin{align}
&\Delta {\bf{H}}_{sr}  \sim \mathcal{CN}_{N_r ,N_s } \left( {{\bf{0}}_{N_r ,N_s } ,{\bf{\Psi }}_{sr}  \otimes {\bf{\Sigma }}_{sr} } \right),\\
&\Delta {\bf{H}}_{rd}  \sim \mathcal{CN}_{N_d ,N_r } \left( {{\bf{0}}_{N_d ,N_r } ,{\bf{\Psi }}_{rd}  \otimes {\bf{\Sigma }}_{rd} } \right),
\end{align}
\noindent with the probability density function (PDF) given by
\begin{align}
&p\left( {\Delta {{\bf{H}}_{sr}}} \right){\rm{ = }}\frac{{\exp \left( {{\rm{ - tr}}\left( {\Delta {\bf{H}}_{sr}^H{\bf{\Sigma }}_{sr}^{{\rm{ - }}1}\Delta {{\bf{H}}_{sr}}{\bf{\Psi }}_{sr}^{{\rm{ - }}1}} \right)} \right)}}{{{{\left( \pi  \right)}^{{N_s}{N_r}}}{{\left| {{{\bf{\Sigma }}_{sr}}} \right|}^{{N_s}}}{{\left| {{{\bf{\Psi }}_{sr}}} \right|}^{{N_r}}}}},\\
&p\left( {\Delta {{\bf{H}}_{rd}}} \right){\rm{ = }}\frac{{\exp \left( {{\rm{ - tr}}\left( {\Delta {\bf{H}}_{rd}^H{\bf{\Sigma }}_{rd}^{{\rm{ - }}1}\Delta {{\bf{H}}_{rd}}{\bf{\Psi }}_{rd}^{{\rm{ - }}1}} \right)} \right)}}{{{{\left( \pi  \right)}^{{N_r}{N_d}}}{{\left| {{{\bf{\Sigma }}_{rd}}} \right|}^{{N_r}}}{{\left| {{{\bf{\Psi }}_{rd}}} \right|}^{{N_d}}}}},
\end{align}

\noindent where ${\bf{\Psi }}_{sr}$ and ${\bf{\Sigma }}_{sr}$ denote the covariance matrices of
the source-to-relay channel seen from the transmitter and receiver, respectively, and so do ${\bf{\Psi }}_{rd}$ and ${\bf{\Sigma }}_{rd}$ for the relay-to-destination channel. The equivalent estimated channel matrix after the ${l_{opt}}$-th transmit ordering pattern can be denoted as ${\bf{\bar H}}_{rd}^{\left( {{l_{opt}}} \right)} = {{\bf{T}}^{\left( {{l_{opt}}} \right)}}{{{\bf{\bar H}}}_{rd}}$.

By using the estimation algorithm proposed in \cite{exponentialmodel}, we have ${{\bf{\Psi }}_{sr}} = {{\bf{R}}_{T,sr}}$, ${{\bf{\Sigma }}_{sr}} = \sigma _{e{\rm{,}}sr}^2{{\bf{R}}_{R,sr}}$, ${{\bf{\Psi }}_{rd}} = {{\bf{R}}_{T,rd}}$ and ${{\bf{\Sigma }}_{rd}} = \sigma _{e{\rm{,}}rd}^2{{\bf{R}}_{R,rd}}$. The matrices ${{\bf{R}}_{T,sr}}$ and ${{\bf{R}}_{R,sr}}$ are the transmit and receive antennas correlation matrices at the source
and the relay, respectively, and $\sigma _{e,sr}^2$ is the source-relay channel estimation error variance. A similar definition can be applied to ${{\bf{R}}_{T,rd}}$, ${{\bf{R}}_{R,rd}}$ and $\sigma _{e,rd}^2$ for the relay-to-destination channel. If we use the channel estimation method proposed in \cite{kronecker}, we then have ${{\bf{\Psi }}_{sr}} = {{\bf{R}}_{T,sr}}$, ${{\bf{\Sigma }}_{sr}} = \sigma _{e{\rm{,}}sr}^2{\left( {{\bf{I}} + \sigma _{e{\rm{,}}sr}^2{\bf{R}}_{R,sr}^{ - 1}} \right)^{{\rm{ - }}1}}$, ${{\bf{\Psi }}_{rd}} = {{\bf{R}}_{T,rd}}$ and ${{\bf{\Sigma }}_{rd}} = \sigma _{e{\rm{,}}rd}^2{\left( {{\bf{I}} + \sigma _{e{\rm{,}}rd}^2{\bf{R}}_{R,rd}^{ - 1}} \right)^{{\rm{ - }}1}}$.  A reasonable assumption that we can make is that ${\bf{\Psi }}_{rd}$, ${\bf{\Sigma }}_{rd}$, ${\bf{\Psi }}_{rd}$ and ${\bf{\Sigma }}_{rd}$ are slowly varying and can be known a priori by estimating long term channel statistics. It is important to note that the analysis to be presented in this paper can be applied in exactly the same way without assuming any specific form of channel estimation error covariance matrix as long as it is symmetric and full-rank\cite{xing2010robust,exponentialmodel,kronecker,biguesh2006training,yoo2004mimo,xing2013general}. It can be shown that if a least squares (LS) channel estimation algorithm
is used to estimate the source-relay and relay-destination channels, the matrices ${\bf{\Psi }}_{sr}$, ${\bf{\Psi }}_{rd}$, ${\bf{\Sigma }}_{sr}$ and ${\bf{\Sigma }}_{rd}$  will be proportional to the identity matrix even for the case where the transmit and receive antennas are correlated\cite{biguesh2006training}.                                             The channel error model we used in this paper is a fairly standard model and widely used for analysis in the existing literature.

\subsection{Problem Formulation}
We focus on the problem of jointly designing ${\bf{F}}_s^{\left( l \right)}$, ${\bf{F}}_r^{\left( l \right)}$, ${{\bf{W}}^{\left( l \right)}}$, ${{\bf{U}}^{\left( l \right)}}$ to minimize the total MSE under the sum power constraint at the source and the relay.
The detailed derivation of the transmit and receive filters is provided in Appendix A.
The system MSE matrix can be written as
\begin{align}
&{\rm{MSE}}\left( {{{\bf{U}}^{\left( l \right)}},{\bf{F}}_s^{\left( l \right)},{\bf{F}}_r^{\left( l \right)},{{\bf{W}}^{\left( l \right)}}} \right) = {\mathop{\rm E}\nolimits} \left[ {{{\left\| {{{\bf{W}}^{\left( l \right)}}{\bf{y}}_d^{\left( l \right)} - {{\bf{v}}^{\left( l \right)}}} \right\|}^2}} \right]\nonumber\\
&  = {\rm{E}}[{\rm{tr}}(\sigma _s^2({{\bf{W}}^{\left( l \right)}}{\bf{H}}_{rd}^{\left( l \right)}{\bf{F}}_r^{\left( l \right)}{{\bf{H}}_{sr}}{\bf{F}}_s^{\left( l \right)} - {{\bf{U}}^{\left( l \right)}})\nonumber\\
& \quad   \times {({{\bf{W}}^{\left( l \right)}}{\bf{H}}_{rd}^{\left( l \right)}{\bf{F}}_r^{\left( l \right)}{{\bf{H}}_{sr}}{\bf{F}}_s^{\left( l \right)} - {{\bf{U}}^{\left( l \right)}})^H})] \nonumber\\
&  \quad    + {\rm{E}}[{\rm{tr}}(\sigma _{{n_{sr}}}^2({{\bf{W}}^{\left( l \right)}}{\bf{H}}_{rd}^{\left( l \right)}{\bf{F}}_r^{\left( l \right)}){({{\bf{W}}^{\left( l \right)}}{\bf{H}}_{rd}^{\left( l \right)}{\bf{F}}_r^{\left( l \right)})^H})]\nonumber\\
&  \quad + {\rm{tr}}(\sigma _{{n_{rd}}}^2{{\bf{W}}^{\left( l \right)}}{{\bf{W}}^{{{\left( l \right)}^H}}}),
\end{align}
\noindent  where the expectation is taken with respect to the channel estimation errors and noise. By taking the expected value, the MSE can be rewritten as
\begin{equation}
\begin{split}
&{\rm{MSE}}\left( {{{\bf{U}}^{\left( l \right)}},{\bf{F}}_s^{\left( l \right)},{\bf{F}}_r^{\left( l \right)},{{\bf{W}}^{\left( l \right)}}} \right) \\
= & {\rm{tr}}\left( {{{\bf{W}}^{\left( l \right)}}{{\bf{A}}^{\left( l \right)}}{{\bf{W}}^{{{\left( l \right)}^H}}}} \right) - \sigma _s^2{\rm{tr}}\left( {{{\bf{U}}^{\left( l \right)}}{\bf{F}}_s^{{{\left( l \right)}^H}}{\bf{\bar H}}_{sr}^H{\bf{F}}_r^{{{\left( l \right)}^H}}{\bf{\bar H}}_{rd}^{{{\left( l \right)}^H}}{{\bf{W}}^{{{\left( l \right)}^H}}}} \right)\\
&  - \sigma _s^2{\rm{tr}}\left( {{{\bf{W}}^{\left( l \right)}}{\bf{\bar H}}_{rd}^{\left( l \right)}{\bf{F}}_r^{\left( l \right)}{{{\bf{\bar H}}}_{sr}}{\bf{F}}_s^{\left( l \right)}{{\bf{U}}^{{{\left( l \right)}^H}}}} \right)+ \sigma _s^2{\rm{tr}}\left( {{{\bf{U}}^{\left( l \right)}}{{\bf{U}}^{{{\left( l \right)}^H}}}} \right),
\end{split}
\end{equation}
\noindent where
\begin{align}
&{{\bf{A}}^{\left( l \right)}} \buildrel \Delta \over = {\bf{\bar H}}_{rd}^{\left( l \right)}{\bf{F}}_r^{\left( l \right)}(\sigma _s^2{{{\bf{\bar H}}}_{sr}}{\bf{F}}_s^{\left( l \right)}{\bf{F}}_s^{{{\left( l \right)}^H}}{\bf{\bar H}}_{sr}^H + \sigma _s^2\alpha _1^{\left( l \right)}{{\bf{\Sigma }}_{sr}}\nonumber\\
&  \quad\quad\quad  + \sigma _{{n_{sr}}}^2{\bf{I}}){\bf{F}}_r^{{{\left( l \right)}^H}}{\bf{\bar H}}_{rd}^{{{\left( l \right)}^H}} + \alpha _2^{\left( l \right)}{{{\bf{\hat \Sigma }}}_{rd}} + \sigma _{{n_{rd}}}^2{\bf{I}}\\
& \alpha _1^{\left( l \right)} \buildrel \Delta \over = {\rm{tr(}}{\bf{F}}_s^{\left( l \right)}{\bf{F}}_s^{{{\left( l \right)}^H}}{{\bf{\Psi }}_{sr}^T}{\rm{)}}\\
&\alpha _2^{\left( l \right)} \buildrel \Delta \over = {\rm{tr}}(({\bf{F}}_r^{\left( l \right)}(\sigma _s^2{{{\bf{\bar H}}}_{sr}}{\bf{F}}_s^{\left( l \right)}{\bf{F}}_s^{{{\left( l \right)}^H}}{\bf{\bar H}}_{sr}^H + \sigma _s^2\alpha _1^{\left( l \right)}{{\bf{\Sigma }}_{sr}}\nonumber\\
& \quad\quad\   + \sigma _{{n_{sr}}}^2{\bf{I}}){\bf{F}}_r^{{{\left( l \right)}^H}}{{\bf{\Psi }}_{rd}^T}))\\
& {{{\bf{\hat \Sigma }}}_{rd}} \buildrel \Delta \over = {{\bf{T}}^{\left( l \right)}}{{\bf{\Sigma }}_{rd}}{{\bf{T}}^{{{\left( l \right)}^H}}}.
\end{align}

By imposing a transmit power constraint at the source and the relay, we arrive at the following
constrained optimization problem:
\begin{align}
&\mathop {\min }\limits_{{{\bf{U}}^{\left( l \right)}},{\bf{F}}_s^{\left( l \right)},{\bf{F}}_r^{\left( l \right)},{{\bf{W}}^{\left( l \right)}}} {\rm{MSE}}\left( {{{\bf{U}}^{\left( l \right)}},{\bf{F}}_s^{\left( l \right)},{\bf{F}}_r^{\left( l \right)},{{\bf{W}}^{\left( l \right)}}} \right)\nonumber\\
& s.t. \quad \ {\rm{tr}}\left( {\sigma _s^2{\bf{F}}_s^{\left( l \right)}{\bf{F}}_s^{{{\left( l \right)}^H}}} \right) \le {P_s} \nonumber\\
& \qquad \ \  {\rm{tr}}\left( {{\bf{F}}_r^{\left( l \right)}\left( {\sigma _s^2{{\bf{H}}_{sr}}{\bf{F}}_s^{\left( l \right)}{\bf{F}}_s^{{{\left( l \right)}^H}}{\bf{H}}_{sr}^H + \sigma _{{n_{sr}}}^2{\bf{I}}} \right){\bf{F}}_r^{{{\left( l \right)}^H}}} \right) \le {P_r},
\end{align}

\section{Proposed Robust Transceiver Design}

In this section, we propose the robust transceiver design for each branch.
The optimal linear receiver ${{\bf{W}}^{\left( l \right)}}$ can be derived by solving $\frac{\partial }{{\partial {{\bf{W}}^{{{\left( l \right)}^*}}}}}{\rm{MSE}}\left( {{{\bf{U}}^{\left( l \right)}},{\bf{F}}_s^{\left( l \right)},{\bf{F}}_r^{\left( l \right)},{{\bf{W}}^{\left( l \right)}}} \right) = {\bf{0}}$, and it is given by
\begin{equation}
{{\bf{W}}^{\left( l \right)}} = \sigma _s^2{{\bf{U}}^{\left( l \right)}}{\bf{F}}_s^{{{\left( l \right)}^H}}{\bf{\bar H}}_{sr}^H{\bf{F}}_r^{{{\left( l \right)}^H}}{\bf{\bar H}}_{rd}^{{{\left( l \right)}^H}}{{\bf{A}}^{{{\left( l \right)}^{{\rm{ - }}1}}}}.
\end{equation}

By substituting (21) into (15) and making use of the matrix inversion lemma \cite{matrixinverse}, the MSE can be expressed as
\begin{equation}
{\rm{MSE}}\left( {{{\bf{U}}^{\left( l \right)}},{\bf{F}}_s^{\left( l \right)},{\bf{F}}_r^{\left( l \right)}} \right) = {\rm{tr}}\left( {{{\bf{E}}^{\left( l \right)}}} \right),
\end{equation}
\noindent where
\begin{align}
& {{\bf{E}}^{\left( l \right)}} \buildrel \Delta \over = {{\bf{U}}^{\left( l \right)}}(\sigma _s^{ - 2}{\bf{I}} + {\bf{F}}_s^{{{\left( l \right)}^H}}{\bf{\bar H}}_{sr}^H{\bf{F}}_r^{{{\left( l \right)}^H}}{\bf{\bar H}}_{rd}^{{{\left( l \right)}^H}}{{\bf{B}}^{{{\left( l \right)}^{ - 1}}}}\nonumber\\
& \quad \quad \quad  \times {\bf{\bar H}}_{rd}^{\left( l \right)}{\bf{F}}_r^{\left( l \right)}{{{\bf{\bar H}}}_{sr}}{\bf{F}}_s^{\left( l \right)}{)^{ - 1}}{{\bf{U}}^{{{\left( l \right)}^H}}}\\
&{{\bf{B}}^{\left( l \right)}} \buildrel \Delta \over = {\bf{\bar H}}_{rd}^{\left( l \right)}{\bf{F}}_r^{\left( l \right)}\left( {\sigma _s^2\alpha _1^{\left( l \right)}{{\bf{\Sigma }}_{sr}} + \sigma _{{n_{sr}}}^2{\bf{I}}} \right){\bf{F}}_r^{{{\left( l \right)}^H}}{\bf{\bar H}}_{rd}^{{{\left( l \right)}^H}}\nonumber\\
& \quad \quad \quad + \alpha _2^{\left( l \right)}{{{\bf{\hat \Sigma }}}_{rd}} + \sigma _{{n_{rd}}}^2{\bf{I}}.
\end{align}

It is well known that for a positive semi-definite matrix ${\bf{M}} \in \mathbb{C}^{N \times N}$, we have $\left| {\bf{M}} \right|^{1/N}  \le {\mathop{\rm tr}\nolimits} \left( {\bf{M}} \right)/N$, which is the arithmetic-geometric mean inequality. Only when ${\bf{M}}$ is a diagonal matrix with equal diagonal elements, the equality can be achieved\cite{THPnodirect}. By letting ${{{\bf{\bar H}}}^{\left( l \right)}} = {\bf{\bar H}}_{rd}^{\left( l \right)}{\bf{F}}_r^{\left( l \right)}{{{\bf{\bar H}}}_{sr}}$, we obtain the following bound on the ${\rm{MSE}}\left( {{{\bf{U}}^{\left( l \right)}},{\bf{F}}_s^{\left( l \right)},{\bf{F}}_r^{\left( l \right)}} \right)$:
\begin{small}
\begin{align}
&\quad \quad \quad \quad \quad{\left| {\left( {\sigma _s^{ - 2}{{\bf{I}}} + {\bf{F}}_s^{{{\left( l \right)}^H}}{{{\bf{\bar H}}}^{{{\left( l \right)}^H}}}{{\bf{B}}^{{{\left( l \right)}^{ - 1}}}}{{{\bf{\bar H}}}^{\left( l \right)}}{\bf{F}}_s^{\left( l \right)}} \right)} \right|^{ - 1/{N_s}}}\nonumber\\
& \le {\rm{tr}}\left\{ {{{\bf{U}}^{\left( l \right)}}\left( {\sigma _s^{ - 2}{{\bf{I}}} + {\bf{F}}_s^{{{\left( l \right)}^H}}{{{\bf{\bar H}}}^{{{\left( l \right)}^H}}}{{\bf{B}}^{{{\left( l \right)}^{ - 1}}}}{{{\bf{\bar H}}}^{\left( l \right)}}{\bf{F}}_s^{\left( l \right)}} \right){{\bf{U}}^{{{\left( l \right)}^H}}}} \right\}/{N_s},
\end{align}
\end{small}
\noindent where the expression of the ${\mathop{\rm MSE}\nolimits}$ in (23) can achieve the lower bound when ${{\bf{E}}^{\left( l \right)}} = \gamma {{\bf{I}}}$,
$\gamma$ is a scaling parameter. Here, we use the fact that $\left|{{\bf{MN}}} \right| = \left| {{\bf{NM}}} \right|$, $\left| {{\bf{M}}^{ - 1} } \right| = \left| {\bf{M}} \right|^{ - 1}$ and $\left| {{\bf{U}}^H {\bf{U}}} \right| = 1$, note that ${\bf{U}}$ is a unit triangular matrix. We also use the rule that for square invertible matrices
 {{\bf{A}} and {{\bf{B}} we have  $\left|{{\bf{AB}}} \right| = \left| {{\bf{A}}} \right|\left| {{\bf{B}}} \right|$}}.
In the following we propose to minimise the lower bound of (25) and find appropriate precoders such that the bound in (25) holds with equality. The constrained optimization problem can be rewritten as
\begin{align}
& \min J\left( {{\bf{F}}_s^{\left( l \right)},{\bf{F}}_r^{\left( l \right)}} \right){\rm{ = }}\left| {{{\left( {\sigma _s^{ - 2}{\bf{I}} + {\bf{F}}_s^{{{\left( l \right)}^H}}{{{\bf{\bar H}}}^{{{\left( l \right)}^H}}}{{\bf{B}}^{{{\left( l \right)}^{ - 1}}}}{{{\bf{\bar H}}}^{\left( l \right)}}{\bf{F}}_s^{\left( l \right)}} \right)}^{{\rm{ - }}1}}} \right|\nonumber\\
& s.t. \quad \ {\rm{tr}}\left( {\sigma _s^2{\bf{F}}_s^{\left( l \right)}{\bf{F}}_s^{{{\left( l \right)}^H}}} \right) \le {P_s} \nonumber\\
& \qquad \ \ {\rm{tr}}({\bf{F}}_r^{\left( l \right)}(\sigma _s^2{{{\bf{\bar H}}}_{sr}}{\bf{F}}_s^{\left( l \right)}{\bf{F}}_s^{{{\left( l \right)}^H}}{\bf{\bar H}}_{sr}^H + \sigma _s^2\alpha _1^{\left( l \right)}{{\bf{\Sigma }}_{sr}}\nonumber \\
& \qquad \ \quad  + \sigma _{{n_{sr}}}^2{{\bf{I}}}){\bf{F}}_r^{{{\left( l \right)}^H}}) \le {P_r},
\end{align}
\noindent We note the fact that the source-relay and relay-destination channels are not completely known. The transmission power consumed by the relay depends on the unknown channel ${{\bf{H}}_{sr}}$.
Thus we take the expectation of the error covariance matrix in formulating the optimisation problem and
Lemma 1 is applied to obtain the expression of the power constraint at the relay.
From (17) and (18), we find that $\alpha _1^{\left( l \right)}$ is a function of ${\bf{F}}_s^{\left( l \right)}$ and $\alpha _2^{\left( l \right)}$ is a function of both ${\bf{F}}_s^{\left( l \right)}$ and ${\bf{F}}_r^{\left( l \right)}$. This problem can be solved by firstly finding the source and relay precoders ${\bf{F}}_s^{\left( l \right)}$ and ${\bf{F}}_r^{\left( l \right)}$ that minimise (26) and satisfy the power constraints, and in a second step using the remaining degrees of freedom to ensure the constraint in (25) holds with equality.

In order to find the explicit structure of the optimal ${{\bf{F}}_s^{\left( l \right)}}$ and ${{\bf{F}}_r^{\left( l \right)}}$, we discuss a scenario with either the covariance matrix of the channel estimation error at the transmitter or a scenario in which the receiver is an identity matrix, respectively. The relations between the scattering environment and the properties of the transmit correlation matrix and the receive correlation matrix are illustrated in \cite{larsson2008space}.  In practice, the transmitter or the receiver is located within a homogenous field of scatterers and we can expect the correlation matrix to be proportional to the identity matrix.  The considered scenarios are represented by the two special cases above.

\subsection{Robust Design with Identity Covariance Matrix at the Transmitter Side}

First of all, we consider the case that the covariance matrix of channel estimation error at the transmitter is an identity matrix, i.e. ${\bf{\Psi }}_{sr} ={\bf{I}}$ and ${\bf{\Psi }}_{rd}  = {\bf{I}}$, we have $\alpha _1^{\left( l \right)} = {\rm{tr}}({\bf{F}}_s^{\left( l \right)}{\bf{F}}_s^{{{\left( l \right)}^H}})$
and $\alpha _2^{\left( l \right)} = {\rm{tr}}({\bf{F}}_r^{\left( l \right)}(\sigma _s^2{{{\bf{\bar H}}}_{sr}}{\bf{F}}_s^{\left( l \right)}{\bf{F}}_s^{{{\left( l \right)}^H}}{\bf{\bar H}}_{sr}^H + \sigma _s^2\alpha _1^{\left( l \right)}{{\bf{\Sigma }}_{sr}} + \sigma _{{n_{sr}}}^2{{\bf{I}}}){\bf{F}}_r^{{{\left( l \right)}^H}})$. It can be shown then that the optimal solution is always achieved with equality in the power constraint.
The optimal solutions of ${{\bf{F}}_s^{\left( l \right)}}$ and ${{\bf{F}}_r^{\left( l \right)}}$ are obtained when $\alpha _1^{\left( l \right)} = {P_s}/\sigma _s^2$ and $\alpha _2^{\left( l \right)} = {P_r}$.
The detailed derivation is provided in Appendix B. The precoding matrices have the following structures:
\begin{align}
&{\bf{F}}_s^{\left( l \right)} = {\bf{\tilde V}}_{sr}^{\left( l \right)}{\bf{\Lambda }}_s^{\left( l \right)}{\bf{\Phi }}_s^{\left( l \right)},\\
& {\bf{\tilde F}}_r^{\left( l \right)} = {\bf{\tilde V }}_{rd}^{\left( l \right)}{\bf{\Lambda }}_r^{\left( l \right)}{\bf{\tilde U}}_{sr}^{{{\left( l \right)}^H}},\\
& {\bf{F}}_r^{\left( l \right)} = {\bf{\tilde F}}_r^{\left( l \right)}{\bf{\tilde \Lambda }}_{{\Sigma _{sr}}}^{{{\left( l \right)}^{{\rm{ - }}\frac{1}{2}}}}{\bf{U}}_{{\Sigma _{sr}}}^H,
\end{align}
\noindent where ${\bf{\Lambda }}_s^{\left( l \right)}$ and ${\bf{\Lambda }}_r^{\left( l \right)}$ are both diagonal matrices with the $i$-th diagonal elements $\lambda _{F_s ,i}$ and $\lambda _{\tilde F_r ,i}$, respectively, and
${\bf{\Phi }}_s^{\left( l \right)}$ is a unitary matrix yet to be determined. Then, we have ${{\bf{B}}^{\left( l \right)}} = {\bf{\bar H}}_{rd}^{\left( l \right)}{\bf{\tilde F}}_r^{\left( l \right)}{\bf{\tilde F}}_r^{{{\left( l \right)}^H}}{\bf{\bar H}}_{rd}^{{{\left( l \right)}^H}} + {\bf{U}}_{{\Sigma _{rd}}}^{\left( l \right)}{\bf{\tilde \Lambda }}_{{\Sigma _{rd}}}^{\left( l \right)}{\bf{U}}_{{\Sigma _{rd}}}^{{{\left( l \right)}^H}}$ . The detailed derivation is shown in Appendix C.

By substituting (27) and (29) into (26), the problem can be simplified as follows:
\begin{align}
&\min J\left( {{\bf{\Lambda }}_s^{\left( l \right)},{\bf{\Lambda }}_r^{\left( l \right)}} \right)\nonumber\\
{\rm{ = }}&\left| {{{\left( {\sigma _s^{ - 2}{\bf{I}} + {\bf{\tilde \Lambda }}_{sr}^{{{\left( l \right)}^2}}{\bf{\Lambda }}_s^{{{\left( l \right)}^2}}{\bf{\Lambda }}_r^{{{\left( l \right)}^2}}{\bf{\tilde \Lambda }}_{rd}^{{{\left( l \right)}^2}}{{\left( {{\bf{\tilde \Lambda }}_{rd}^{{{\left( l \right)}^2}}{\bf{\Lambda }}_r^{{{\left( l \right)}^2}} + {\bf{I}}} \right)}^{ - 1}}} \right)}^{{\rm{ - }}1}}} \right|\nonumber\\
& s.t.\quad {\rm{tr}}\left( {\sigma _s^2{\bf{\Lambda }}_s^{{{\left( l \right)}^2}}} \right) \le {P_s}\nonumber\\
& \qquad \  {\rm{tr}}\left( {{\bf{\Lambda }}_r^{{{\left( l \right)}^2}}\left( {\sigma _s^2{\bf{\Lambda }}_s^{{{\left( l \right)}^2}}{\bf{\tilde \Lambda }}_{sr}^{{{\left( l \right)}^2}} + {{\bf{I}}}} \right)} \right) \le {P_r},
\end{align}
Note that for a positive semi-definite matrix ${\bf{M}} \in \mathbb{C}^{N \times N}$, we have \cite{matrixinverse}
\begin{equation}
\det \left( {\bf{M}} \right) \le \prod\limits_{i = 1}^{N } {{\bf{M}}\left( {i,i} \right)},
\end{equation}
\noindent the equality holds when ${\bf{M}}$ is a diagonal matrix\cite{diagonal}.

Let $\tilde \lambda _{1,i}$ and $\tilde \lambda _{2,i}$ be the $i$th diagonal element of ${\bf{\tilde \Lambda }}_{sr}^{\left( l \right)}$ and ${\bf{\tilde \Lambda }}_{rd}^{\left( l \right)}$, respectively, $i{\rm{ = }}1{\rm{,}} \cdots {\rm{,}}{N_d}$, from (30) we have the
following results
\begin{align}
&\mathop {\min }\limits_{{\lambda _{{F_s},i}},{\lambda _{{{\tilde F}_r},i}}} \prod\limits_{i = 1}^{{N_d}} {{{\left( {\sigma _s^{ - 2} + \frac{{\tilde \lambda _{1,i}^2\tilde \lambda _{2,i}^2\lambda _{{F_s},i}^2\lambda _{{{\tilde F}_r},i}^2}}{{\tilde \lambda _{2,i}^2\lambda _{{{\tilde F}_r},i}^2 + 1}}} \right)}^{{\rm{ - }}1}}}\\
& s.t.\quad \sum\limits_{i = 1}^{N_d } {\sigma _s^2 \lambda _{F_s ,i}^2  \le P_s }\\
& \qquad \  \sum\limits_{i = 1}^{N_d } {\lambda _{\tilde F_r ,i}^2 \left( {\sigma _s^2 \lambda _{F_s ,i}^2 \tilde \lambda _{1,i}^2  + 1} \right)}  \le P_r.
\end{align}

Let us introduce
\begin{align}
& x_i  \buildrel \Delta \over = \sigma _s^2 \lambda _{F_s ,i}^2\\
& y_i  \buildrel \Delta \over = \lambda _{\tilde F_r ,i}^2 \left( {\sigma _s^2 \lambda _{F_s ,i}^2
 \tilde \lambda _{1,i}^2  + 1} \right),
\end{align}
\noindent (32) becomes a maximization problem and the logarithm is used. This is possible because the logarithmic function is a monotonically increasing function which makes the formulated problem equivalent. Thus
the optimization problem can be rewritten as
\begin{align}
& \max \sum\limits_{i = 1}^{N_d } {\ln \left( {\frac{{y_i \tilde \lambda _{2,i}^2 x_i \tilde
\lambda _{1,i}^2  + y_i \tilde \lambda _{2,i}^2  + x_i \tilde \lambda _{1,i}^2  + 1}}{{y_i \tilde
\lambda _{2,i}^2  + x_i \tilde \lambda _{1,i}^2  + 1}}} \right)}\\
& s.t.\quad \sum\limits_{i = 1}^{N_d } {x_i }  \le P_s\nonumber\\
& \qquad \  \sum\limits_{i = 1}^{N_d } {y_i }  \le P_r.
\end{align}
 The solution to the objective function can be obtained by using an iterative waterfilling method\cite{unified} via KKT conditions\cite{boyd2004convex}.  The detailed derivation is summarized in Appendix D.
 For a given $x_i$, by solving (37) and (38), the optimum $y_i$ can be obtained as follows:
\begin{equation}
y_i  = \frac{1}{{2\tilde \lambda _{2,i}^2 }}\left[ {\sqrt {\tilde \lambda _{1,i}^4 x_i^2  + 4\tilde
\lambda _{1,i}^2 x_i \tilde \lambda _{2,i}^2 \mu _r }  - \tilde \lambda _{1,i}^2 x_i  - 2} \right]^ +,
\end{equation}

\noindent where $\left[ y \right]^ +   = \max \left[ {0,y} \right]$, and $\mu _r$ is the water level which satisfies the power constraint with equality at the relay in (38). By solving (37) and (38), the optimum $x_i$ can be calculated as
\begin{equation}
x_i  = \frac{1}{{2\tilde \lambda _{1,i}^2 }}\left[ {\sqrt {\tilde \lambda _{2,i}^4 y_i^2  + 4\tilde
 \lambda _{1,i}^2 y_i \tilde \lambda _{2,i}^2 \mu _s }  - \tilde \lambda _{2,i}^2 y_i  - 2} \right]^ +,
\end{equation}

\noindent where $\mu _s$ is the water level which satisfies the power constraint with equality at the source in (38). The algorithm can be implemented iteratively with initial values.
 Note that $\lambda _{F_s ,i}$ and $\lambda _{\tilde F_r ,i}$ can be calculated based on (35) and (36).
 Note that this iterative water-filling algorithm is
guaranteed to converge, as discussed in \cite{yu2004iterative}.
As shown in \cite{unified}, a locally optimal solution can be obtained
by iteratively updating the power allocation vector of one node by fixing the power allocation vectors of all other nodes.
 We then focus on the derivation of the unitary matrix ${\bf{\Phi }}_s^{\left( l \right)}$ and the feedback matrix ${{{\bf{U}}^{\left( l \right)}}}$.

 The lower bound of MSE is achieved when the objective function in (23) is a diagonal matrix with equal diagonal elements. Thus, the following equation must be satisfied:
 \begin{equation}
 {{\bf{U}}^{\left( l \right)}}\left( {\sigma _s^{ - 2}{{\bf{I}}} + {\bf{F}}_s^{{{\left( l \right)}^H}}{{{\bf{\bar H}}}^{{{\left( l \right)}^H}}}{{\bf{B}}^{{{\left( l \right)}^{ - 1}}}}{{{\bf{\bar H}}}^{\left( l \right)}}{\bf{F}}_s^{\left( l \right)}} \right){{\bf{U}}^{{{\left( l \right)}^H}}} = {{\bar \sigma }^2}{{\bf{I}}}.
 \end{equation}

  By substituting (27) and (29) into (41), we obtain ${{\bf{U}}^{\left( l \right)}}{\bf{\Phi }}_s^{{{\left( l \right)}^H}}{{\bf{\Sigma }}^{{{\left( l \right)}^{ - 1/2}}}}{{\bf{\Sigma }}^{{{\left( l \right)}^{ - 1/2}}}}{\bf{\Phi }}_s^{\left( l \right)}{{\bf{U}}^{{{\left( l \right)}^H}}} = {{\bar \sigma }^2}{{\bf{I}}}$. Then we define ${{{\bf{\tilde U}}}^{\left( l \right)}}{\rm{ = }}\bar \sigma {{\bf{U}}^{{{\left( l \right)}^{{\rm{ - }}H}}}}$ and apply the geometric mean decomposition (GMD)\cite{gmd} to ${{\bf{\Sigma }}^{{{\left( l \right)}^{ - 1/2}}}}$ to make the diagonal entries of an upper triangular matrix all equal. In\cite{gmd}, the GMD was proved to be asymptotically optimal for high SNR, in terms of both channel throughput and BER performance. Thus, we obtain ${{\bf{\Sigma }}^{{{\left( l \right)}^{ - 1/2}}}} = {{\bf{Q}}^{\left( l \right)}}{{{\bf{\tilde U}}}^{\left( l \right)}}{\bf{\Phi }}_s^{{{\left( l \right)}^H}}$,  where ${{\bf{Q}}^{\left( l \right)}}$ and ${\bf{\Phi }}_s^{\left( l \right)}$ are unitary matrices, and ${{{\bf{\tilde U}}}^{\left( l \right)}}$ is an upper triangular matrix with equal diagonal elements $\bar \sigma$, where $\bar \sigma^2$ is given by
\begin{equation}
\bar \sigma ^2  = \prod\limits_{i = 1}^{N_s } {\left( {\sigma _s^{ - 2}  + \frac{{\tilde \lambda _{1,i}^2 \tilde \lambda _{2,i}^2 \lambda _{F_s ,i}^2 \lambda _{\tilde F_r ,i}^2 }}{{\tilde \lambda _{2,i}^2 \lambda _{\tilde F_r ,i}^2  + 1}}} \right)} ^{ - 1/N_s },
\end{equation}
\\[-5mm]
From the equation above, it can be verified that the equality is achieved. We then calculate  ${{\bf{U}}^{\left( l \right)}} = \bar \sigma {{{\bf{\tilde U}}}^{{{\left( l \right)}^{{\rm{ - }}H}}}}$. With ${\bf{\Phi }}_s^{\left( l \right)}$ and ${{\bf{U}}^{\left( l \right)}}$, the source and relay precoders corresponding to the $l$-th cancellation order are obtained by (27) and (29). Subsequently, the MMSE receiver ${{{\bf{W}}^{\left( l \right)}}}$ can be derived by substituting (27) and (29) into (21).

\subsection{Robust Design with Identity Covariance Matrix at the Receiver Side}
Then, we consider the case that the covariance matrix of the channel estimation error at the receiver side is an identity matrix, i.e., ${\bf{\Sigma }}_{sr}  = \sigma _e^2 {\bf{I}}$ and ${\bf{\Sigma }}_{rd}  = \sigma _e^2 {\bf{I}}$.
We perform the SVD of the estimated channels:
$ {{{\bf{\bar H}}}_{sr}} = {{\bf{U}}_{sr}}{{\bf{\Lambda }}_{sr}}{\bf{V}}_{sr}^H$,
$ {\bf{\bar H}}_{rd}^{\left( l \right)} = {\bf{U}}_{rd}^{\left( l \right)}{\bf{\Lambda }}_{rd}^{\left( l \right)}{\bf{V}}_{rd}^{{{\left( l \right)}^H}}$,
it can be clearly seen from (76)-(81) that for ${\bf{\Sigma }}_{sr}  = \sigma _e^2 {\bf{I}}$ and ${\bf{\Sigma }}_{rd}  = \sigma _e^2 {\bf{I}}$, we have ${{{\bf{\tilde U}}}_{sr}}{\rm{ = }}{{\bf{U}}_{sr}}$, ${\bf{\tilde U}}_{rd}^{\left( l \right)}{\rm{ = }}{\bf{U}}_{rd}^{\left( l \right)}$, ${{{\bf{\tilde V}}}_{sr}}{\rm{ = }}{{\bf{V}}_{sr}}$ and ${\bf{\tilde V}}_{rd}^{\left( l \right)}{\rm{ = }}{\bf{V}}_{rd}^{\left( l \right)}$.

Thus, the precoding matrices have the following structure:
\begin{align}
& {\bf{F}}_s^{\left( l \right)} = {{\bf{V}}_{sr}}{\bf{\Lambda }}_s^{\left( l \right)}{\bf{\Phi }}_s^{\left( l \right)},\\
& {\bf{F}}_r^{\left( l \right)} = {\bf{V}}_{rd}^{\left( l \right)}{\bf{\Lambda }}_r^{\left( l \right)}{\bf{U}}_{sr}^H,\\
& {{\bf{B}}^{\left( l \right)}} = {\beta _1}{\bf{\bar H}}_{rd}^{\left( l \right)}{\bf{F}}_r^{\left( l \right)}{\bf{F}}_r^{{{\left( l \right)}^H}}{\bf{\bar H}}_{rd}^{{{\left( l \right)}^H}} + {\beta _2}{{\bf{I}}},
\end{align}
\noindent where
\begin{align}
& {\beta _1} = \sigma _e^2\cdot\sigma _s^2{\rm{tr}}\left( {{\bf{F}}_s^{\left( l \right)}{\bf{F}}_s^{{{\left( l \right)}^H}}{{\bf{\Psi }}_{sr}^T}} \right) + \sigma _{{n_{sr}}}^2,\\
& {\beta _2} = \sigma _e^2{\rm{tr\{ }}{\bf{F}}_r^{\left( l \right)}{\rm{(}}\sigma _s^2{{{\bf{\bar H}}}_{sr}}{\bf{F}}_s^{\left( l \right)}{\bf{F}}_s^{{{\left( l \right)}^H}}{\bf{\bar H}}_{sr}^H + {\beta _1}{{\bf{I}}}{\rm{)}}{\bf{F}}_r^{{{\left( l \right)}^H}}{{\bf{\Psi }}_{rd}^T}{\rm{\} }} + \sigma _{{n_{rd}}}^2.
\end{align}

The constrained optimization that corresponds to the proposed robust design can be written as
\begin{align}
 &\min {\prod\limits_{i = 1}^{{N_d}} {\left( {\sigma _s^{ - 2} + \frac{{\lambda _{1,i}^2\lambda _{2,i}^2\lambda _{{F_s},i}^2\lambda _{{F_r},i}^2}}{{{\beta _1}\lambda _{2,i}^2\lambda _{{F_r},i}^2 + {\beta _2}}}} \right)} ^{{\rm{ - }}1}},\\
& s.t.\quad \sum\limits_{i = 1}^{N_d } {\sigma _s^2 \lambda _{F_s ,i}^2  \le P_s },\\
& \qquad \  \sum\limits_{i = 1}^{N_d } {\lambda _{F_r ,i}^2 \left( {\sigma _s^2 \lambda _{F_s ,i}^2
 \lambda _{1,i}^2  + \beta _1 } \right)}  \le P_r.
\end{align}

By introducing the following definitions
\begin{align}
& x_i  \buildrel \Delta \over = \sigma _s^2 \lambda _{F_s ,i}^2,\\
& y_i  \buildrel \Delta \over = \lambda _{F_r ,i}^2 \left( {\sigma _s^2 \lambda _{F_s ,i}^2
\lambda _{1,i}^2  + \beta _1 } \right),
\end{align}

The optimization can be finally formulated as
\begin{align}
& \max \sum\limits_{i = 1}^{N_d } {\ln \left( {\frac{{y_i \lambda _{2,i}^2 x_i \lambda _{1,i}^2  + y_i
 \lambda _{2,i}^2 \beta _1  + x_i \lambda _{1,i}^2 \beta _2  + \beta _1 \beta _2 }}{{y_i
 \lambda _{2,i}^2 \beta _1  + x_i \lambda _{1,i}^2 \beta _2  + \beta _1 \beta _2 }}} \right)},\\
& s.t.\quad \sum\limits_{i = 1}^{N_d } {x_i }  \le P_s,\\
& \qquad \  \sum\limits_{i = 1}^{N_d } {y_i }  \le P_r,
\end{align}
\noindent where $\lambda _{1,i}$ and $\lambda _{2,i}$ are the $i$th diagonal elements of ${\bf{\Lambda }}_{sr}^{\left( l \right)}$ and $
{\bf{\Lambda }}_{rd}^{\left( l \right)}$, respectively. The quantities $\beta _1$ and $\beta _2$ can be written as
\begin{align}
& \beta _1  \buildrel \Delta \over = \sum\limits_{i = 1}^{N_s } {\sigma _e^2 x_i \left[ {{\bf{\tilde
\Psi }}_{sr} } \right]_{ii} }  + \sigma _{n_{sr} }^2,\\
& \beta _2  \buildrel \Delta \over = \sum\limits_{i = 1}^{N_s } {\sigma _e^2 y_i \left[ {{\bf{\tilde
 \Psi }}_{rd} } \right]_{ii}  + } \sigma _{n_{rd} }^2,\\
 & {\bf{\tilde \Psi }}_{sr}  \buildrel \Delta \over = {\bf{V}}_{sr}^H {\bf{\Psi }}_{sr}^T {\bf{V}}_{sr},\\
& {{\bf{\tilde \Psi }}_{rd}} \buildrel \Delta \over = {\bf{V}}_{rd}^{{{{\rm{(}}l{\rm{)}}}^H}}{{\bf{\Psi }}_{rd}^T}{\bf{V}}_{rd}^{{\rm{(}}l{\rm{)}}}.
\end{align}
 We find the following solutions to the optimization problem by using the
aforementioned iterative method:
\begin{align}
& x_i  = \frac{{\beta _1 }}{{2\lambda _{1,i}^2 }}\left[ {\sqrt {\frac{{\lambda _{2,i}^4 y_i^2 }}
{{\beta _2^2 }} + \frac{{4\lambda _{1,i}^2 y_i \lambda _{2,i}^2 \mu _s }}{{\beta _1 \beta _2 }}}
- \frac{{\lambda _{2,i}^2 y_i }}{{\beta _2 }} - 2} \right]^ +,\\
& y_i  = \frac{{\beta _2 }}{{2\lambda _{2,i}^2 }}\left[ {\sqrt {\frac{{\lambda _{1,i}^4 x_i^2 }}
{{\beta _1^2 }} + \frac{{4\lambda _{1,i}^2 x_i \lambda _{2,i}^2 \mu _r }}{{\beta _1 \beta _2 }}}
 - \frac{{\lambda _{1,i}^2 x_i }}{{\beta _1 }} - 2} \right]^ +.
\end{align}

Similarly, we can derive the values of ${\bf{\Phi }}_s^{\left( l \right)}$ and  ${{{\bf{U}}^{\left( l \right)}}}$. Finally, the source and relay precoders can be obtained explicitly.
In Section V, we will show the simulation results of the proposed robust THP scheme.

\section{Selection Criterion, Complexity Analysis and Ordering Schemes}

We have presented the overall principles and structures of the proposed algorithm in the previous section. In this section, we introduce the selection criterion, the complexity analysis and the ordering schemes which are employed in our proposed design.

\subsection{Selection Criterion for the Proposed MB-THP Scheme}

A proper selection criterion is of great importance for the MB-THP algorithm to achieve a significant performance improvement in MIMO relay systems. We have investigated a number of different criteria and the squared Euclidean distance has been identified as a simple and yet effective selection mechanism.
 The selection criterion chooses the best branch corresponding to the minimum squared Euclidean distance, which is given by
\begin{equation}
{l_{{\rm{opt}}}}{\rm{ = }}\arg \mathop {\min }\limits_{1 \le l \le L} {{\bf{J}}^{\left( l \right)}}\left( i \right),
\end{equation}
\noindent where ${{\bf{J}}^{\left( l \right)}}\left( i \right)$ is the squared Euclidean distance corresponding to the $l$-th cancellation branch for the $i$-th transmission data block, which is expressed by
\begin{equation}
{{\bf{J}}^{\left( l \right)}}\left( i \right) = {\left\| {{\bf{b}}\left( i \right) - {{{\bf{\hat b}}}^{(l)}}\left( i \right)} \right\|^2},
\end{equation}
\noindent where
 the quantity ${\bf{b}}\left( i \right)$ denotes the $i$-th transmission data block, which is given by ${\bf{b}}\left( i \right) = {\left[ {{{\bf{s}}^T}\left( i \right),...,{{\bf{s}}^T}\left( {i + K - 1} \right)} \right]^T}$,
$K$ is the block length, the vector ${\bf{s}}\left( {i + k} \right)$ denotes the $k$-th transmit vector of the $i$-th block, $k \in \left\{ {1,...,K - 1} \right\}$.
 ${{{\bf{\hat b}}}^{(l)}}\left( i \right)$ is the transformed version of ${{{\bf{\tilde b}}}^{(l)}}\left( i \right)$  back to the original order for the $i$-th block,
 and the vector ${{{\bf{\tilde b}}}^{(l)}}\left( i \right)$ denotes the noise-free pre-estimated values of the data at the transmitter using estimated CSI, which is given by
 \begin{equation}
 {{{\bf{\tilde b}}}^{(l)}}\left( i \right){\rm{ = MOD}}\left( {{{{\bf{\tilde r}}}^{(l)}}\left( i \right)} \right),
 \end{equation}
 \noindent where ${{{\bf{\tilde r}}}^{(l)}}\left( i \right) = {\left[ {{{{\bf{\tilde y}}}^{{{\left( l \right)}^T}}}\left( i \right),...,{{{\bf{\tilde y}}}^{{{\left( l \right)}^T}}}\left( {i + K - 1} \right)} \right]^T}$.  ${{{\bf{\tilde y}}}^{\left( l \right)}}\left( {i + k} \right)$ denotes the pre-estimated received vector based on the $l$-th branch for the $k$-th transmit vector of the $i$-th block, which is expressed as follows
 \begin{equation}
 {{{\bf{\tilde y}}}^{\left( l \right)}}\left( {i + k} \right){\rm{ = }}{{\bf{W}}^{\left( l \right)}}{\bf{\bar H}}_{rd}^{\left( l \right)}{\bf{F}}_r^{\left( l \right)}{{{\bf{\bar H}}}_{sr}}{\bf{F}}_s^{\left( l \right)}{{{\bf{\bar x}}}^{\left( l \right)}}\left( {i + k} \right).
 \end{equation}

 The optimum branch is chosen by minimizing the summation of the squared Euclidean distance values in one transmission data block. The selected optimum branch is updated once per block. Ideally, the optimum branch can be chosen to minimize the accumulated squared Euclidean distance between the
true transmit symbol and the received soft information at the destination in one transmission block.
However, the selection criterion is conducted at the transmitter, we cannot obtain the exact received signal at the destination. We then adopt the noiseless information to estimate the received signal in our proposed algorithm.   The simulation results in Section V show that the proposed scheme achieves a better performance with respect to the conventional
algorithms, which verifies the effectiveness of the approximation. It is worth to mention that since the proposed algorithm is implemented at the transmitter, the selection criterion has the full information of the transmit symbols.
The procedure of the proposed robust transceiver algorithm for the case that the covariance matrix of channel estimation error at the transmitter is an identity matrix is summarized in Table I.

 \begin{table*}[t]
\centering
 \caption{\normalsize {Proposed Robust Transceiver Algorithm}} {
\begin{tabular}{ll}
\hline
\hline
 $\textsc{1}$ & $\mathbf{for}$ each parallel branch $l$, $l \in \left\{ {1,...,L} \right\}$.\\
$\textsc{2}$ & ~~Solve for the unknown diagonal matrices ${\bf{\Lambda }}_s^{\left( l \right)}$ and ${\bf{\Lambda }}_r^{\left( l \right)}$ in the optimal precoding structure by selecting an appropriate initial  \\
& ~~choice for $x_i$ that satisfies$\sum\limits_{i = 1}^{{N_d}} {{x_i}} {\rm{ = }}{P_s}$, the algorithm updates $y_i$ according to
$y_i  = \frac{1}{{2\tilde \lambda _{2,i}^2 }}\left[ {\sqrt {\tilde \lambda _{1,i}^4 x_i^2  + 4\tilde
\lambda _{1,i}^2 x_i \tilde \lambda _{2,i}^2 \mu _r }  - \tilde \lambda _{1,i}^2 x_i  - 2} \right]^ +$ \\
& ~~and $x_i$ according to $x_i  = \frac{1}{{2\tilde \lambda _{1,i}^2 }}\left[ {\sqrt {\tilde \lambda _{2,i}^4 y_i^2  + 4\tilde\lambda _{1,i}^2 y_i \tilde \lambda _{2,i}^2 \mu _s }  - \tilde \lambda _{2,i}^2 y_i  - 2} \right]^ +$ in an alternating way, note that the variable $\mu _r$ and  \\
& ~~$\mu _r$ can be solved by the bisection method detailed in\cite{THPdirect}.\\
  $\textsc{3}$ & ~~Compute ${\bf{\Phi }}_s^{\left( l \right)}$ and the feedback matrix ${{\bf{U}}^{\left( l \right)}}$ based on ${{\bf{U}}^{\left( l \right)}}\left( {\sigma _s^{ - 2}{{\bf{I}}} + {\bf{F}}_s^{{{\left( l \right)}^H}}{{{\bf{\bar H}}}^{{{\left( l \right)}^H}}}{{\bf{B}}^{{{\left( l \right)}^{ - 1}}}}{{{\bf{\bar H}}}^{\left( l \right)}}{\bf{F}}_s^{\left( l \right)}} \right){{\bf{U}}^{{{\left( l \right)}^H}}} = {{\bar \sigma }^2}{{\bf{I}}}$,\\
     & ~~${\bf{F}}_s^{\left( l \right)} = {\bf{\tilde V}}_{sr}^{\left( l \right)}{\bf{\Lambda }}_s^{\left( l \right)}{\bf{\Phi }}_s^{\left( l \right)}$, and apply the GMD to ${{\bf{\Sigma }}^{{{\left( l \right)}^{ - 1/2}}}}$, ${{\bf{\Sigma }}^{{{\left( l \right)}^{ - 1/2}}}} = {{\bf{Q}}^{\left( l \right)}}{{{\bf{\tilde U}}}^{\left( l \right)}}{\bf{\Phi }}_s^{{{\left( l \right)}^H}}$.  \\
 $\textsc{4}$ & ~~Derive the optimal structure of ${\bf{F}}_s^{\left( l \right)}$ and ${\bf{F}}_r^{\left( l \right)}$ given by ${\bf{F}}_s^{\left( l \right)} = {\bf{\tilde V}}_{sr}^{\left( l \right)}{\bf{\Lambda }}_s^{\left( l \right)}{\bf{\Phi }}_s^{\left( l \right)}$ and ${\bf{F}}_r^{\left( l \right)} = {\bf{\tilde F}}_r^{\left( l \right)}{\bf{\tilde \Lambda }}_{{\Sigma _{sr}}}^{{{\left( l \right)}^{{\rm{ - }}\frac{1}{2}}}}{\bf{U}}_{{\Sigma _{sr}}}^H$.\\
 $\textsc{5}$ & ~~Compute the receiver ${{\bf{W}}^{\left( l \right)}}$ by using the obtained ${\bf{F}}_s^{\left( l \right)}$, ${\bf{F}}_r^{\left( l \right)}$ and ${{\bf{U}}^{\left( l \right)}}$.\\
 $\textsc{6}$ & ~~Compute the squared Euclidean distance for the $l$-th
  cancellation order, ${{\bf{J}}^{\left( l \right)}}\left( i \right) = {\left\| {{\bf{b}}\left( i \right) - {{{\bf{\hat b}}}^{(l)}}\left( i \right)} \right\|^2}$.\\
 $\textsc{7}$ & $\mathbf{end}$\\
 $\textsc{8}$ & Choose the optimum branch by using the selection criterion
   for data transmission and send the optimum index to the relay \\
  & and receiver with the aid of the limited feedforward transmission.\\
 \hline
\label{tab:table1}
\end{tabular}
}
\end{table*}
\subsection{Complexity Analysis}

In this part, we focus on the  computational complexity of the proposed robust precoding scheme.
Complexity is measured in terms of the number of FLOPs, defined as the floating-point operations. Note that from \cite{matrixcomputations}, we know that a complex addition and multiplication have 2 and 6 FLOPs, respectively. The computational complexity of the proposed scheme is summarized in Table II. As we can see, the proposed methods mainly involve singular value decomposition (SVD), matrix multiplications, matrix inversions, and the GMD. Essentially, we detail the complexity of the proposed robust MB-THP procedure in each branch, and the overall complexity of the MB-THP algorithm can be obtained by multiplying the complexity
of the robust THP per branch by the number of branches $L$. Also, the selection mechanism requires $O(KN_d^2)$ operations. The computational complexity of the
regular TH joint source and linear relay precoding algorithm in \cite{THPnodirect} is
$O(N_s^3 + {N_r}N_s^2 + {N_d}N_s^2 + {N_d}N_r^2 + N_r^3 + {N_s}N_r^2 + N_d^3 + {N_s}N_d^2{\rm{ + }}{N_d}{I_s}{I_i}{\rm{ + }}{N_d}{I_r}{I_i})$. We advocate an affordable increase in complexity in exchange for the improvement of the performance.
The proposed scheme can effectively improve the performance including the ability to
mitigate the multistream interference, alleviate the effects of CSI errors and enhance the robustness of the system.

\begin{table*}[t]
\centering
\caption{\normalsize {
Complexity of Proposed MB-THP Algorithm in Each Branch}} {
\begin{tabular}{|c|c|c|}\hline
\hline\rule{0cm}{2.0ex}
Step & Operation & FLOPs    \\\hline
\ $\textsc{1}$  &  ${\bf{\tilde H}}_{sr}^{\left( l \right)}$  & $O{\rm{(}}N_s^2{\rm{(}}{N_s}{\rm{ + }}{N_r}{\rm{ + }}{N_d}{\rm{))}}$    \\\hline\rule{0cm}{2.0ex}
$\textsc{2}$  &   ${\bf{\tilde H}}_{rd}^{\left( l \right)}$  &  $O{\rm{(}}N_r^2{\rm{(}}{N_s}{\rm{ + }}{N_r}{\rm{ + }}{N_d}{\rm{))}}$     \\\hline\rule{0cm}{2.0ex}
$\textsc{3}$  &   $\textmd{SVD}\ \ {\bf{\tilde H}}_{sr}^{\left( l \right)} = {\bf{\tilde U}}_{sr}^{\left( l \right)}{\bf{\tilde \Lambda }}_{sr}^{\left( l \right)}{\bf{\tilde V}}_{sr}^{{{\left( l \right)}^H}}$  &  $O{\rm{(}}{N_r}N_s^2{\rm{ + }}N_s^3{\rm{)}}$     \\\hline\rule{0cm}{2.0ex}
$\textsc{4}$ &   $\textmd{SVD}\ \ {\bf{\tilde H}}_{rd}^{\left( l \right)} = {\bf{\tilde U}}_{rd}^{\left( l \right)}{\bf{\tilde \Lambda }}_{rd}^{\left( l \right)}{\bf{\tilde V}}_{rd}^{{{\left( l \right)}^H}}$  &  $O{\rm{(}}{N_d}N_r^2{\rm{ + }}N_r^3{\rm{)}}$     \\\hline\rule{0cm}{2.0ex}
$\textsc{5}$  &   ${{\bf{B}}^{{{\left( l \right)}^{ - 1}}}}$  &  $O{\rm{(}}N_d^3{\rm{)}}$     \\\hline\rule{0cm}{2.0ex}
$\textsc{6}$  &   ${x_i} \ \textmd{and} \ {y_i}$  &  $O({N_d} {I_s}{I_i}{\rm{ + }}{N_d} {I_r}{I_i})$     \\\hline\rule{0cm}{3.0ex}
$\textsc{7}$  &   $\textmd{GMD}\ \ {{\bf{\Sigma }}^{{{\left( l \right)}^{ - 1/2}}}} = {{\bf{Q}}^{\left( l \right)}}{{{\bf{\tilde U}}}^{\left( l \right)}}{\bf{\Phi }}_s^{{{\left( l \right)}^H}}$  &  $O(N_d^3{\rm{)}}$     \\\hline
\ $\textsc{8}$  &   ${{\bf{U}}^{\left( l \right)}} = \bar \sigma {{{\bf{\tilde U}}}^{{{\left( l \right)}^{{\rm{ - }}H}}}}$  &  $O(N_d^3{\rm{)}}$     \\\hline
\ $\textsc{9}$  &   ${{\bf{F}}_s} \ \textmd{and} \ {{\bf{F}}_r}$   &  $O(({N_s}{N_d}{\rm{ + }}{N_s}N_d^2){\rm{ + }}(N_r^2{\rm{ + }}N_r^3))$    \\\hline
\multicolumn{3}{|l|}{${I_r}: \textmd{number of iteration for computing} \ {\bf{\Lambda }}_r^{\left( l \right)} $}\\
\multicolumn{3}{|l|}{${I_s}: \textmd{number of iteration for computing} \ {\bf{\Lambda }}_s^{\left( l \right)} $}\\
\multicolumn{3}{|l|}{${I_i}: \textmd{number of iteration of the water-filling process} $}\\\hline
 \end{tabular}
}
\end{table*}

\subsection{Ordering Schemes}

A V-BLAST like ordering strategy for THP has been studied in \cite{windpassinger2003precoding}  .
The V-BLAST ordering algorithm requires multiple calculations of the pseudo inverse of the channel matrix.
Therefore, a suboptimal heuristic sorted LQ decomposition algorithm has been  extended from the sorted QR decomposition (SQRD) algorithm
in \cite{wubben2002efficient}. And a tree search (TS) algorithm has
also been proposed in \cite{habendorf2006ordering}.
 The above ordering algorithms, however, assume that each distributed receiver is equipped with a single antenna. The cooperative ordering processing is impractical for MIMO relay systems.
 And the TS algorithm shows superior performance especially for medium to high SNRs.
In order to achieve a better BER performance in the whole SNR range, we proposed the MB strategy.
It is clear that the optimal ordering scheme which conducts an exhaustive
search is complex for practical systems, especially when the number of transmit antennas is large. Thus, we discuss two sub-optimal off-line schemes to design the transformation matrices ${{\bf{T}}^{\left( l \right)}}$ with appropriate structures such that they can be used for low-complexity implementation. The basic arbitrary ordering scheme randomly chooses a subset from the exhaustive search. However, the proposed schemes are developed to select a subset from the optimal ordering scheme set in a way that approaches the optimal performance while keeping the computational complexity low.

1) Pre-Stored Patterns (PSP): The transformation matrix ${{\bf{T}}^{{\rm{(1)}}}}$ for the first branch is chosen as the identity matrix ${\bf{I}}$ to keep the optimal ordering as described by
${{\bf{T}}^{{\rm{(1)}}}} = {\bf{I}}$. The remaining ordering patterns can be described mathematically by
\begin{equation}
{{\bf{T}}^{\left( l \right)}}{\rm{ = }}\left[ {\begin{array}{*{20}{c}}
{{{\bf{I}}_s}}&{{{\bf{0}}_{s{\rm{,}}{N_s}{\rm{ - }}s}}}\\
{{{\bf{0}}_{{N_s}{\rm{ - }}s{\rm{,}}s}}}&{\phi \left[ {{{\bf{I}}_s}} \right]}
\end{array}} \right]{\rm{, }}  \ 2 \le l \le {N_s},
\end{equation}
where ${{\bf{0}}_{m{\rm{,}}n}}$ denotes an $m \times n$-dimensional matrix full of zeros, the operator $\phi \left[  \cdot  \right]$ rotates the elements of the argument matrix column-wise such that an identity matrix becomes a matrix with ones in the reverse diagonal. The proposed ordering algorithm shifts the ordering of the cancellation according to shifts given by
\begin{equation}
s{\rm{ = }}\left\lfloor {\left( {l{\rm{ - }}2} \right){N_s}{\rm{/}}L} \right\rfloor ,\;2 \le l \le {N_s},
\end{equation}
\noindent where $L$ is the number of parallel branches. In order to
illustrate this problem clearly, we take the situation when the number of branches $L=4$ as an example. By using the scheme above, we obtain the transformation matrix ${{\bf{T}}^{(l)}}$ as follows
\begin{equation*}
{{\bf{T}}^{\left( {\rm{1}} \right)}}{\rm{ = }}\left[ {\begin{array}{*{20}{c}}
{\rm{1}}&{\rm{0}}&{\rm{0}}&{\rm{0}}\\
{\rm{0}}&{\rm{1}}&{\rm{0}}&{\rm{0}}\\
{\rm{0}}&{\rm{0}}&{\rm{1}}&{\rm{0}}\\
{\rm{0}}&{\rm{0}}&{\rm{0}}&{\rm{1}}
\end{array}} \right]{\rm{,}}\ {{\bf{T}}^{\left( {\rm{2}} \right)}}{\rm{ = }}\left[ {\begin{array}{*{20}{c}}
{\rm{0}}&{\rm{0}}&{\rm{0}}&{\rm{1}}\\
{\rm{0}}&{\rm{0}}&{\rm{1}}&{\rm{0}}\\
{\rm{0}}&{\rm{1}}&{\rm{0}}&{\rm{0}}\\
{\rm{1}}&{\rm{0}}&{\rm{0}}&{\rm{0}}
\end{array}} \right],
\end{equation*}

\begin{equation}
{{\bf{T}}^{\left( 3 \right)}}{\rm{ = }}\left[ {\begin{array}{*{20}{c}}
{\rm{1}}&{\rm{0}}&{\rm{0}}&{\rm{0}}\\
{\rm{0}}&0&{\rm{0}}&1\\
{\rm{0}}&{\rm{0}}&{\rm{1}}&{\rm{0}}\\
{\rm{0}}&1&{\rm{0}}&0
\end{array}} \right]{\rm{,}}\ {{\bf{T}}^{\left( 4 \right)}}{\rm{ = }}\left[ {\begin{array}{*{20}{c}}
{\rm{1}}&{\rm{0}}&{\rm{0}}&{\rm{0}}\\
{\rm{0}}&{\rm{1}}&{\rm{0}}&{\rm{0}}\\
{\rm{0}}&{\rm{0}}&0&1\\
{\rm{0}}&{\rm{0}}&1&0
\end{array}} \right].
\end{equation}

2) Frequently Selected Branches (FSB): The FSB algorithm builds a codebook which contains the ordering patterns for the most likely selected branches and the required number of branches to obtain a near-optimal performance is greatly reduced. In order to build the FSB codebook, we need to perform an extensive set of experiments and compute the frequency
of the indices of the selected patterns to identify the statistics of each selected branch and construct the codebook with the $L$ most likely selected branches to be encountered. The algorithm is summarized in Table III, where ${{\bf{d}}_E}$ denotes the vector of Euclidean distance for all possible branches, ${N_e}$ denotes the total number of experiments we did, ${{\bf{L}}_{idx}}$ is defined for the storage of the selected branches for every experiment and ${{\bf{L}}_o}$ is the codebook for optimal ordering patterns computed by ${\bf{PERMS}}({N_s}:{\rm{ - }}1:1)$, which provides the list containing all possible permutations of the ${N_s}$ elements. We highlight that in each run, after we measure the Euclidean distances for all branches, the branch that results in the smallest Euclidean distance is stored in ${{\bf{L}}_{idx}}$ at step 10. Finally, the FSB codebook ${{\bf{L}}_{FSB}}$ is created by selecting the most frequently selected L branches according to the histogram of ${{\bf{L}}_{idx}}$.
 \begin{table}[h]
\centering
 \caption{\normalsize {Frequently Selected Branches Ordering Scheme}} {
\begin{tabular}{cll}
\hline
\hline
 $\textsc{1}$ & ${{\bf{d}}_E} \leftarrow {\bf{NULL}}$,\ ${{\bf{L}}_{idx}} \leftarrow {\bf{NULL}}$, \ ${{\bf{L}}_{FSB}} \leftarrow {\bf{NULL}}$\\
 $\textsc{2}$ & ${L_{opt}} \leftarrow {N_s}!$,\ $l \leftarrow 1$ \\
 $\textsc{3}$ & ${{\bf{L}}_o} \leftarrow {\bf{PERMS}}{\rm{(}}{N_s}{\rm{: - }}1{\rm{:}}\ 1{\rm{)}}$\\
 $\textsc{4}$ & $\mathbf{for}$ ${n_e}{\rm{ = }}1\ \textmd{to} \ {N_e}\  \mathbf{do}$\\
 $\textsc{5}$ & ~~$\mathbf{for}$ ${l}{\rm{ = }}1\ \textmd{to} \ {{L_{opt}}}\  \mathbf{do}$\\
 $\textsc{6}$ & ~~~~${{\bf{T}}^{\left( l \right)}} \leftarrow {{\bf{L}}_o}{\rm{(}}l{\rm{)}}$ \\
 $\textsc{7}$ & ~~~~${{{\bf{\hat s}}}^{(l)}} \leftarrow {\bf{SIC}}({{\bf{T}}^{\left( l \right)}}{\bf{\bar {\bf H}}})$ \\
 $\textsc{8}$ & ~~~~${{\bf{d}}_E}\left( l \right) \leftarrow \left\| {{\bf{s}} - {{{\bf{\hat s}}}^{(l)}}} \right\|$ \\
 $\textsc{9}$ & ~~$\mathbf{end \ for}$\\
 $\textsc{10}$ & ~~${{\bf{L}}_{idx}}{\rm{(}}{n_e}{\rm{)}} \leftarrow {\bf{MIN\_Index}}{\rm{(}}{{\bf{d}}_E}{\rm{)}}$  \\
 $\textsc{11}$ & $\mathbf{end \ for}$\\
 $\textsc{12}$ & ${{\bf{L}}_{FSB}} \leftarrow {\bf{SELECT}}({\bf{HIST}}({{\bf{L}}_{idx}}))$\\
\hline
\label{tab:table1}
\end{tabular}
}
\end{table}
\subsection{Efficiency}

From the algorithm discussed above, we know that for every block prior to the data transmission, the source sends the index of the selected optimal branch which is chosen by the selection rule to the relay and the destination through a limited feedforward channel. We insert the limited feedforward bits at the beginning of the corresponding transmission block. Each transmission block comprises $K$ symbol periods each one consisting of ${N_d}$ spatial streams, and the feedforward rate of the optimum index is one per transmission block.
We consider a $m$-ary modulation and assume that $B$ bits index information to be sent for each transmit block. Thus, the transmission efficiency is given by
\begin{equation}
\varepsilon {\rm{ = }}\frac{{{N_d}K{{\log }_2}\left( m \right)}}{{{N_d}K{{\log }_2}\left( m \right) + B}}.
\end{equation}

In this work, we employ the 16-QAM modulation and employ a data block of $K = 100$ symbols in the simulation. For the exhaustive search ordering scheme with 24 branches, we need 5 feedforward bits. For a configuration with $N_s  = N_r  = N_d  = 4$,
by using $B = 5$ feedforward bits we achieve the transmission efficiency of 99.68\%.
 For a slow fading channel, the feedforward rate is very low, and the transmission efficiency is close to 1. It should be noted
that the efficiency can be made higher if we increase the block length $K$. In the simulation section, we will show that with the side information (SI) the performance of the proposed precoding algorithm outperforms the performance of the conventional precoding algorithms significantly.

\section{Simulation Results}

In this section, we evaluate the performance of the proposed precoding scheme. We adopt a simulation approach and conduct several experiments in order to verify the effectiveness of the proposed techniques. In the following, we consider an AF MIMO relay system with $N_s  = N_r  = N_d  = 4$. By using the exponential model \cite{kronecker}, the channel estimation error covariance matrices can be expressed as
\begin{equation}
{\bf{\Psi }}_{sr}  = {\bf{\Psi }}_{rd}  = \begin{bmatrix}
1 & \alpha & \alpha ^2 & \alpha ^3\\
\alpha & 1 & \alpha  & \alpha ^2\\
\alpha ^2 & \alpha  & 1 & \alpha\\
\alpha ^3 & \alpha ^2 & \alpha & 1,\nonumber
\end{bmatrix}
\end{equation}

\begin{equation}
{\bf{\Sigma }}_{sr}  = {\bf{\Sigma }}_{rd}  =\sigma _e^2 \begin{bmatrix}
1 & \beta & \beta ^2 & \beta ^3\\
\beta & 1 & \beta  & \beta ^2\\
\beta ^2 & \beta  & 1 & \beta\\
\beta ^3 & \beta ^2 & \beta & 1
\end{bmatrix},
\end{equation}

\noindent where $\alpha$ and $\beta$ denote the correlation coefficients, and $\sigma _e^2$ is the estimation error variance. The estimated channels, ${\bf{\bar H}}_{sr}$ and
${\bf{\bar H}}_{rd}$, are generated by the distributions as follows:
\begin{equation}
 {\bf{\bar H}}_{sr} \sim\mathcal{CN}_{N_r ,N_s } \left( {{\bf{0}}_{N_r ,N_s } ,\frac{{\left( {1 - \sigma _e^2 } \right)}}{{\sigma _e^2 }}{\bf{\Psi }}_{sr} } \otimes {\bf{\Sigma }}_{sr}  \right)
\end{equation}
\begin{equation}
{\bf{\bar H}}_{rd} \sim \mathcal{CN}_{N_d ,N_r } \left( {{\bf{0}}_{N_d ,N_r } ,\frac{{\left( {1 - \sigma _e^2 } \right)}}{{\sigma _e^2 }} {\bf{\Psi }}_{rd} } \otimes {\bf{\Sigma }}_{rd} \right),
\end{equation}
\noindent such that channel realizations have unit variance. In the simulation, for the data transmission process the SNR at the relay is defined as ${\mathop{\rm SNR}\nolimits} _{sr} = {P_s}/\sigma _{{n_{sr}}}^2$, and the SNR at the destination is defined as ${\mathop{\rm SNR}\nolimits} _{rd} = {P_r}/\sigma _{{n_{rd}}}^2$.
 We adopt the diagonal elements of the identity matrix as the initial values for the iterative algorithm.  Also, we use 16-QAM as the modulation scheme.

\begin{figure}
\centering
\includegraphics[scale=.47]{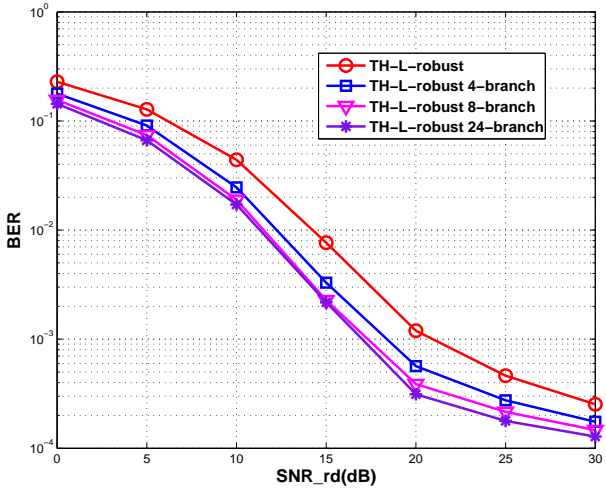}
\caption{\small{BER performance comparison for different MB ordering schemes. ($\alpha=\beta =0, {\sigma _e^2 }=0.001$)
}}
\end{figure}

\begin{figure}\centering
\includegraphics[scale=.48]{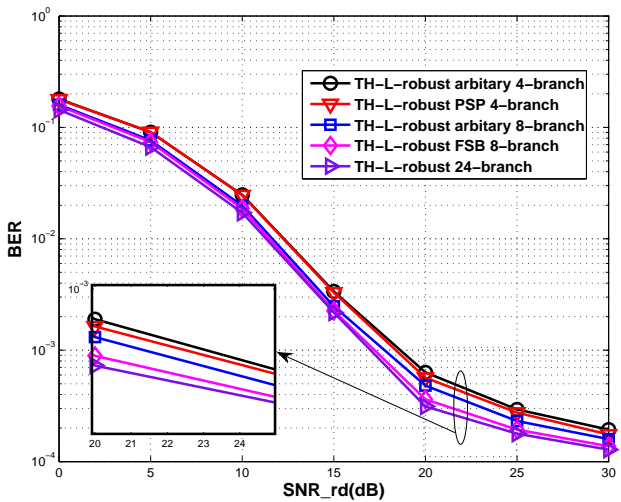}
\caption{\small{BER performance comparison for pre-designed and arbitrary MB ordering schemes. ($\alpha=\beta =0, {\sigma _e^2 }=0.001$)}}
\end{figure}

\begin{figure}\centering
\includegraphics[scale=.5]{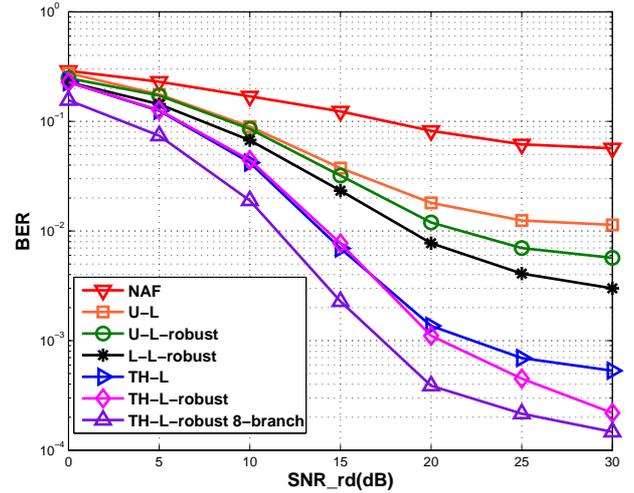}
\caption{\small{BER performance comparison for conventional precoding techniques and the proposed MB-THP algorithm.
($\alpha=\beta =0, {\sigma _e^2 }=0.001$)}}
\end{figure}

\begin{figure}\centering
\includegraphics[scale=.52]{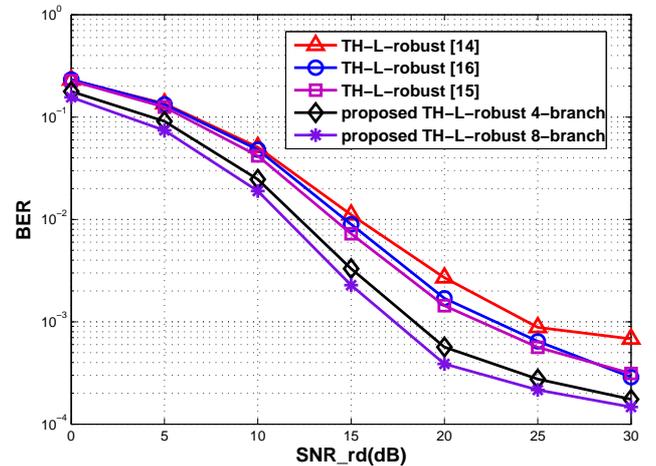}
\caption{\small{BER performance comparison for existing robust THP techniques and the proposed MB-THP algorithm.
($\alpha=\beta =0, {\sigma _e^2 }=0.001$)}}
\end{figure}

\begin{figure}
\centering
\includegraphics[scale=.48]{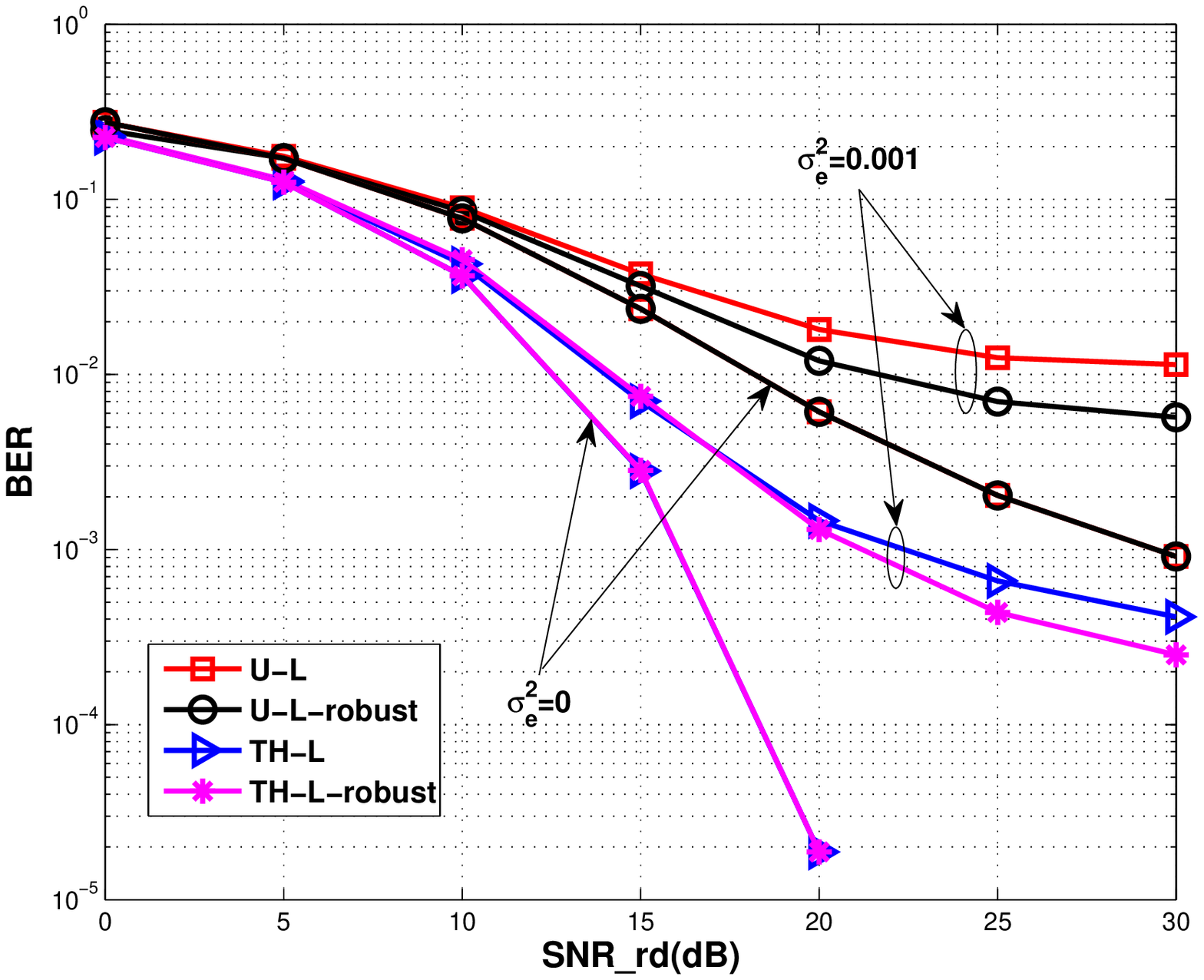}
\caption{\small{BER performance comparison for existing precoding algorithms and the proposed robust THP.
($\alpha=\beta =0, {\sigma _e^2 }=0\setminus{\sigma _e^2 }=0.001$)
}}
\end{figure}

\begin{figure}
\centering
\includegraphics[scale=.50]{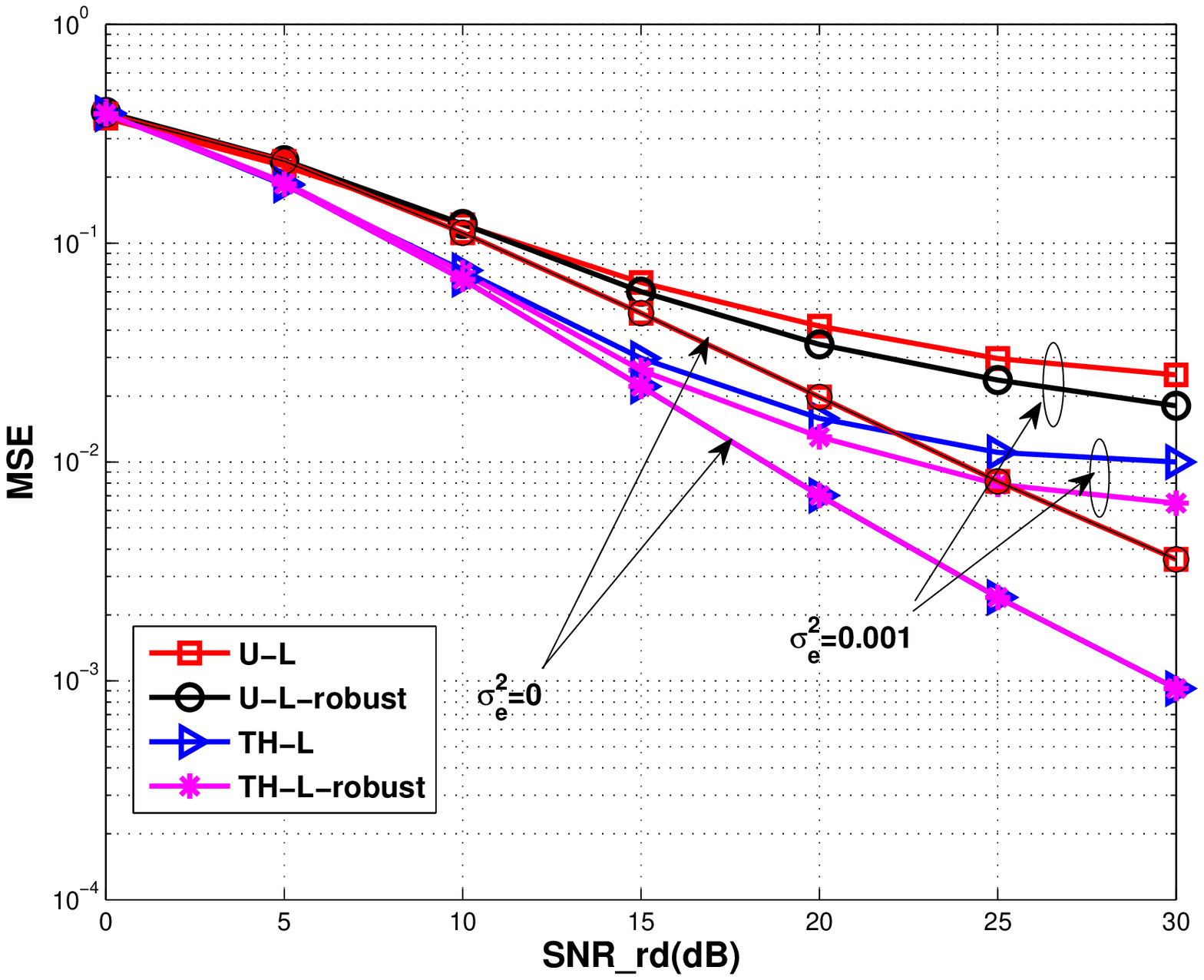}
\caption{\small{MSE performance comparison for existing precoding algorithms and the proposed robust THP.
($\alpha=\beta =0, {\sigma _e^2 }=0\setminus{\sigma _e^2 }=0.001$)
}}
\end{figure}

\begin{figure}
\centering
\includegraphics[scale=.53]{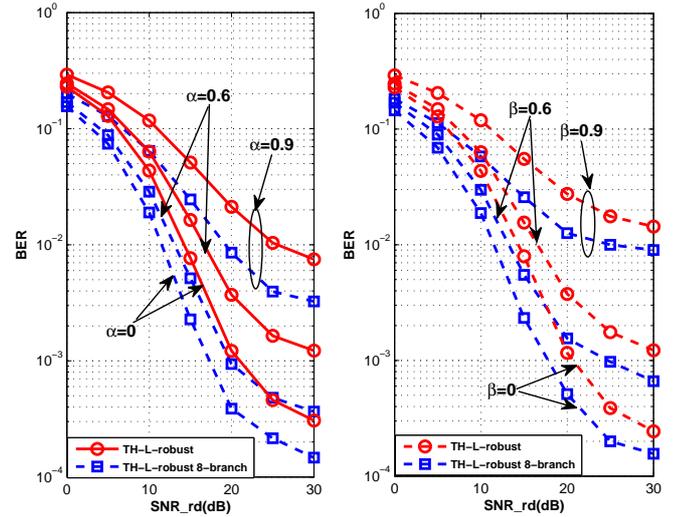}
\caption{\small{ BER performance comparison for the robust THP system and the proposed MB-THP algorithm
with different $\alpha$  and $\beta$.}}
\end{figure}

\begin{figure}
\centering
\includegraphics[scale=.49]{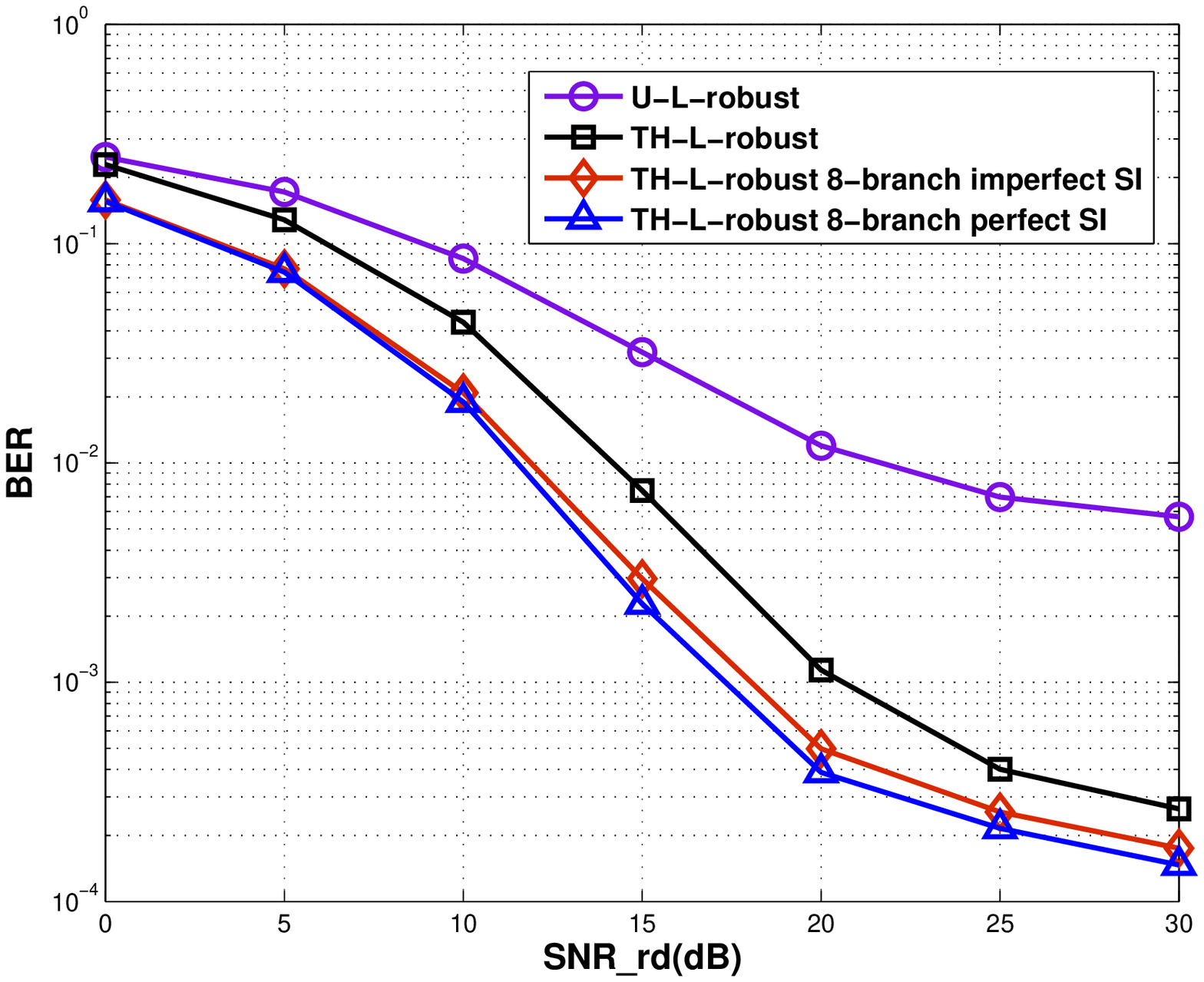}
\caption{\small{BER performance comparison for the robust THP system and the proposed MB-THP algorithm
with perfect and imperfect SI.
}}
\end{figure}

According to the above settings, we use a Monte-Carlo approach to obtain the required expected values over numerous channel realizations. Here, we let ${\mathop{\rm SNR}\nolimits} _{sr}=30\ {\mathop{\rm dB}\nolimits}$, $\alpha=\beta =0, {\sigma _e^2 }=0.001$ and ${\mathop{\rm SNR}\nolimits} _{rd}$ is varied.
Fig. 2 shows the BER performance versus the SNR for comparing the proposed MB-THP transceiver scheme, i.e., 4-, 8-, 24- pre-designed cancellation ordering branches, respectively.
The proposed robust TH source and linear relay
precoded system without considering the ordering schemes (TH-L-robust) is also listed here for comparison.
The best performance is achieved with the proposed scheme with 24 ordering branches, i.e., the exhaustive search. The BER decreases as the number of branches increases. The plots also show that the performance of the robust FSB algorithm with 8 branches approaches the optimal ordering scheme closely.

We then compare the BER performance for the pre-designed and the arbitrary MB ordering schemes of the proposed MB-THP algorithm under the scenario that $\alpha=\beta =0$ and ${\sigma _e^2 }=0.001$. As shown in Fig. 3, the measurements agree quite well with the simulations. By employing the pre-designed MB ordering scheme described in Section IV, the BER performance can be further improved for $L=4$ and $L=8$ branches, respectively. Moreover, the performance of each pre-designed MB ordering scheme is always superior to the arbitrary ordering scheme, respectively. Accordingly, we only consider the sub-optimal FSB algorithm of the MB ordering schemes in the following simulations for low-complexity implementation.

In the third set of simulations, we also let ${\mathop{\rm SNR}\nolimits} _{sr}=30\ {\mathop{\rm dB}\nolimits}$, $\alpha=\beta =0, {\sigma _e^2 }=0.001$ and ${\mathop{\rm SNR}\nolimits} _{rd}$ is varied. Here, we compare the proposed robust MB-THP algorithm with the following five existing MIMO relay precoding algorithms and the proposed robust THP algorithm without considering the ordering schemes:
1) a non-precoded system with a Wiener filter (NAF);  2) the linear relay precoded system without source precoding (U-L)\cite{relayreceiver}; 3) the robust linear relay precoded system without source precoding (U-L-robust)\cite{xing2010robust};  4) the linear robust joint source and relay precoded system (L-L-robust)\cite{jointrobust};
5) the TH source and linear relay precoded system (TH-L)\cite{THPnodirect}; 6) the proposed robust TH source and linear relay
precoded system (TH-L-robust). As shown in Fig. 4, the proposed robust MB-THP algorithm using the sub-optimal FSB algorithm with $L=8$ branches outperforms the existing transceiver designs in terms of BER. Meanwhile, the performance of the proposed robust algorithm considering the estimation error is better than that of the conventional non-robust algorithms estimate the channels directly. Specifically, the proposed robust MB-THP algorithm can lead to 3 dB gain in SNR in comparison
with the proposed robust THP algorithm without considering the ordering schemes (TH-L-robust), and can save up to almost 5 dB in comparison
with the conventional TH source and linear relay precoding algorithm (TH-L) , at the BER level of ${\rm{1}}{{\rm{0}}^{{\rm{ - 3}}}}$.

Fig. 5 shows the BER comparison for our proposed MB-THP algorithm with the recently mentioned robust algorithms
which consist of THP at the source along with a linear precoder at the relay
in \cite{tseng2012robust,xing2012robust,millar2012robust}.
Note that the robust algorithm in \cite{xing2012robust} considers the multi-hop relay system. For fair comparison, here we focus on the algorithm for two-hop system only.
As expected, the proposed method outperforms all the other algorithms. A degree of performance improvement
is achieved compared to the recently described robust algorithms.
The proposed robust MB-THP algorithm with $L=8$ branches can lead to 3.5 dB gain in SNR in comparison
with the robust THP algorithm in\cite{xing2012robust}, and can save 4 dB and 6
dB, compared with the
the precoding algorithms in\cite{tseng2012robust} and \cite{millar2012robust}, respectively, at the BER level of ${\rm{1}}{{\rm{0}}^{{\rm{ - 3}}}}$.

Next, we investigate the effect of the channel estimation errors on the BER and MSE performances. Also let ${{\mathop{\rm SNR}\nolimits} _{sr}}{\rm{ = }}30{\mathop{\rm dB}\nolimits}$, $\alpha=\beta =0, \sigma _e^2 = 0\ {\rm{(}}\sigma _e^2 = 0.001{\rm{)}}$, and ${{\mathop{\rm SNR}\nolimits} _{rd}}$ is varied. Here, we further incorporate the linear relay precoded system without source precoding (U-L)\cite{relayreceiver} and the robust linear relay precoded system without source precoding (U-L-robust)\cite{xing2010robust} for comparison. From Fig. 6, we observe that, as expected, since U-L-robust only considers a relay precoder, its performance is inferior to the proposed
TH-L-robust no matter if CSI is perfect or not. On the other hand, the performance of the algorithm based on
estimated channels only shows performance degradation compared to that of the two robust algorithms in terms of imperfect CSI. As expected, the performance of the corresponding robust and nonrobust algorithms coincide when $\sigma _e^2 = 0$ for perfect CSI. Fig. 7 shows the corresponding MSE performance, which is consistent with the BER performance.

Fig. 8 shows the BER performance comparison for the proposed robust THP system and the proposed MB-THP algorithm with different values of the correlation coefficients. For the left figure, we let $\beta=0, {\sigma _e^2 }=0.001$ and $\alpha$ is varied. It can be seen that smaller correlation coefficients lead to a better performance. When the value of $\alpha$ decreases, the performance of both algorithms improves. Of course, the performance of the proposed MB-THP algorithm is always superior to the performance of the proposed robust THP (TH-L-robust) algorithm. In particular, the proposed MB-THP algorithm can save up to almost 4 dB in comparison with the algorithm without ordering scheme, at the BER level of ${\rm{1}}{{\rm{0}}^{{\rm{ - 3}}}}$ when $\alpha=0$.
 Furthermore, the performance gap between the TH-L-robust and MB-THP becomes larger as
$\alpha$ increases. The right figure shows the BER performance comparison for the proposed robust THP system and the proposed MB-THP algorithm with different values of $\beta$.  Here, we let $\alpha=0, {\sigma _e^2 }=0.001$ and $\beta$ is varied. It can be seen that a similar conclusion can be drawn. Those curves saturate in the high SNR region.

The results in Fig. 9 show the BER performance versus ${\rm{SN}}{{\rm{R}}_{rd}}$ for the proposed robust MB-THP preprocessing scheme
and the conventional linear precoding system using perfect and imperfect SI at the transmitter. We use a structure based on a frame format where the indices are converted to 0s and 1s. This frame of 1s and 0s with the feedforward information is transmitted over a binary symmetric channel with an associated probability of error. We assume that there is a 1\% SI error of the optimal index information, which corresponds to almost 1 dB degradation, compared with the perfect SI case at a BER level of ${\rm{1}}{{\rm{0}}^{{\rm{ - 3}}}}$. This case shows the ability of our proposed algorithms to deal with SI errors. In order to make sure the SI error are controlled, channel coding techniques can be applied to the signalling feedforward channels with errors.

\section{Conclusion}
In this paper, a robust MB-THP transceiver design in MIMO relay networks with imperfect CSI has been proposed. The proposed MB structure is equipped with several parallel branches based on pre-designed ordering patterns. For each branch, the transceiver is composed of a TH precoder at the source, a linear precoder at the relay and an MMSE receiver at the destination. The solution for the precoders has been finally obtained by using an iterative method via the KKT conditions. An appropriate selection rule has been developed to choose the nonlinear transceiver corresponding to the best branch for data transmission. Simulations have shown that the proposed robust design outperforms the existing non-precoded/precoded systems without taking the channel uncertainties into account.

\appendices
\section{Derivation of the transmit and receive filters }

\begin{lemma}
For a random matrix ${\bf{M}} \in \mathbb{C}^{M \times N}$ with a multi-variate Gaussian distribution
${\bf{A}}\sim \mathcal{CN}\left( {{\bf{\bar A}},{\bf{C}} \otimes {\bf{D}}} \right)$, we have for any
matrix ${\bf{F}} \in \mathbb{C}^{N \times N}$ that ${\mathop{\rm E}\nolimits} [{\bf{AF}}{{\bf{A}}^H}] = {\bf{\bar AF}}{{{\bf{\bar A}}}^H} + {\mathop{\rm tr}\nolimits} {\rm{(}}{\bf{F}}{{\bf{C}}^T}{\rm{)}}{\bf{D}}$\cite{matrixdistribution}.
\end{lemma}

Using Lemma 1, we have
\begin{align}
{\rm{E}}[{{\bf{H}}_{sr}}{\bf{F}}_s^{\left( l \right)}{\bf{F}}_s^{{{\left( l \right)}^H}}{\bf{H}}_{sr}^H]
=& {\rm{E}}[{\rm{(}}{{{\bf{\bar H}}}_{sr}}{\rm{ + }}\Delta {{\bf{H}}_{sr}}{\rm{)}}{\bf{F}}_s^{\left( l \right)}{\bf{F}}_s^{{{\left( l \right)}^H}}{{\rm{(}}{{{\bf{\bar H}}}_{sr}}{\rm{ + }}\Delta {{\bf{H}}_{sr}}{\rm{)}}^H}]\nonumber\\
=&{{\bf{\bar H}}_{sr}}{\bf{F}}_s^{\left( l \right)}{\bf{F}}_s^{{{\left( l \right)}^H}}{\bf{\bar H}}_{sr}^H + {\rm{tr}}({\bf{F}}_s^{\left( l \right)}{\bf{F}}_s^{{{\left( l \right)}^H}}{{\bf{\Psi }}_{sr}^T}){{\bf{\Sigma }}_{sr}},
\end{align}
\noindent where ${\bf{\Psi }}_{sr}$ and ${\bf{\Sigma }}_{sr}$ denote the covariance matrices of
the source-to-relay channel seen from the transmitter and the receiver, respectively.

Similarly, we obtain
\begin{align}
&{\rm{E}}[{\bf{H}}_{rd}^{\left( l \right)}{\bf{F}}_r^{\left( l \right)}{{\bf{H}}_{sr}}{\bf{F}}_s^{\left( l \right)}{\bf{F}}_s^{{{\left( l \right)}^H}}{\bf{H}}_{sr}^H{\bf{F}}_r^{{{\left( l \right)}^H}}{\bf{H}}_{rd}^{{{\left( l \right)}^H}}{\rm{]}} \nonumber\\
=& {\rm{E}}[{\bf{H}}_{rd}^{\left( l \right)}{\bf{F}}_r^{\left( l \right)}({{{\bf{\bar H}}}_{sr}}{\bf{F}}_s^{\left( l \right)}{\bf{F}}_s^{{{\left( l \right)}^H}}{\bf{\bar H}}_{sr}^H\nonumber\\
& \quad + {\rm{tr}}({\bf{F}}_s^{\left( l \right)}{\bf{F}}_s^{{{\left( l \right)}^H}}{{\bf{\Psi }}_{sr}^T}){{\bf{\Sigma }}_{sr}}){\bf{F}}_r^{{{\left( l \right)}^H}}{\bf{H}}_{rd}^{{{\left( l \right)}^H}}{\rm{]}}\nonumber\\
=&{\bf{\bar H}}_{rd}^{\left( l \right)}{\bf{F}}_r^{\left( l \right)}({{{\bf{\bar H}}}_{sr}}{\bf{F}}_s^{\left( l \right)}{\bf{F}}_s^{{{\left( l \right)}^H}}{\bf{\bar H}}_{sr}^H\nonumber\\
& + {\rm{tr}}({\bf{F}}_s^{\left( l \right)}{\bf{F}}_s^{{{\left( l \right)}^H}}{{\bf{\Psi }}_{sr}^T}){{\bf{\Sigma }}_{sr}}){\bf{F}}_r^{{{\left( l \right)}^H}}{\bf{\bar H}}_{rd}^{{{\left( l \right)}^H}}\nonumber\\
&{\rm{ + tr}}({\bf{F}}_r^{\left( l \right)}({{{\bf{\bar H}}}_{sr}}{\bf{F}}_s^{\left( l \right)}{\bf{F}}_s^{{{\left( l \right)}^H}}{\bf{\bar H}}_{sr}^H\nonumber\\
& + {\rm{tr}}({\bf{F}}_s^{\left( l \right)}{\bf{F}}_s^{{{\left( l \right)}^H}}{{\bf{\Psi }}_{sr}^T}){{\bf{\Sigma }}_{sr}}){\bf{F}}_r^{{{\left( l \right)}^H}}{{\bf{\Psi }}_{rd}^T}){{{\bf{\hat \Sigma }}}_{rd}}.
\end{align}

We also have
\begin{equation}
{\rm{E}}[{\bf{H}}_{rd}^{\left( l \right)}{\bf{F}}_r^{\left( l \right)}{{\bf{H}}_{sr}}{\bf{F}}_s^{\left( l \right)}]{\rm{ = }}{\bf{\bar H}}_{rd}^{\left( l \right)}{\bf{F}}_r^{\left( l \right)}{{{\bf{\bar H}}}_{sr}}{\bf{F}}_s^{\left( l \right)}.
\end{equation}

\section{Proof of Conditions for the Optimal Solution }
By introducing ${\bf{\tilde F}}_s^{\left( l \right)}{\rm{ = }}{{{\bf{F}}_s^{\left( l \right)}} \mathord{\left/
 {\vphantom {{{\bf{F}}_s^{\left( l \right)}} {\sqrt {{\rm{tr}}({\bf{F}}_s^{\left( l \right)}{\bf{F}}_s^{{{\left( l \right)}^H}})} }}} \right.
 \kern-\nulldelimiterspace} {\sqrt {{\rm{tr}}({\bf{F}}_s^{\left( l \right)}{\bf{F}}_s^{{{\left( l \right)}^H}})} }}$, we can rewrite the MSE as
\begin{equation}
{\rm{ |}}(\sigma _s^{ - 2}{{\bf{I}}} + {\bf{\tilde F}}_s^{{{\left( l \right)}^H}}{\bf{\bar H}}_{sr}^H{\bf{F}}_r^{{{\left( l \right)}^H}}{\bf{\bar H}}_{rd}^{{{\left( l \right)}^H}}{{{\bf{\bar B}}}^{{{\left( l \right)}^{ - 1}}}}{\bf{\bar H}}_{rd}^{\left( l \right)}{\bf{F}}_r^{\left( l \right)}{{{\bf{\bar H}}}_{sr}}{\bf{\tilde F}}_s^{\left( l \right)}){\rm{|}}\nonumber,
\end{equation}

\noindent where
\begin{align}
&{{{\bf{\bar B}}}^{\left( l \right)}} \buildrel \Delta \over = {\bf{\bar H}}_{rd}^{\left( l \right)}{\bf{F}}_r^{\left( l \right)}( {\sigma _s^2{{\bf{\Sigma }}_{sr}} + {{\sigma _{{n_{sr}}}^2{{\bf{I}}}} \mathord{\left/
 {\vphantom {{\sigma _{{n_{sr}}}^2{{\bf{I}}_{{N_r}}}} {{\rm{tr}}({\bf{F}}_s^{\left( l \right)}{\bf{F}}_s^{{{\left( l \right)}^H}})}}} \right.
 \kern-\nulldelimiterspace} {{\rm{tr}}({\bf{F}}_s^{\left( l \right)}{\bf{F}}_s^{{{\left( l \right)}^H}})}}} ){\bf{F}}_r^{{{\left( l \right)}^H}}{\bf{\bar H}}_{rd}^{{{\left( l \right)}^H}}\nonumber\\
& \quad \quad \quad + \tilde \alpha _2^{\left( l \right)}{{{\bf{\hat \Sigma }}}_{rd}} + {{\sigma _{{n_{rd}}}^2{{\bf{I}}}} \mathord{\left/
 {\vphantom {{\sigma _{{n_{rd}}}^2{{\bf{I}}}} {{\rm{tr}}({\bf{F}}_s^{\left( l \right)}{\bf{F}}_s^{{{\left( l \right)}^H}})}}} \right.
 \kern-\nulldelimiterspace} {{\rm{tr}}({\bf{F}}_s^{\left( l \right)}{\bf{F}}_s^{{{\left( l \right)}^H}})}}\nonumber,\\
&\tilde \alpha _2^{\left( l \right)} \buildrel \Delta \over = {\rm{tr}}({\bf{F}}_r^{\left( l \right)}(\sigma _s^2{{{\bf{\bar H}}}_{sr}}{\bf{\tilde F}}_s^{\left( l \right)}{\bf{\tilde F}}_s^{{{\left( l \right)}^H}}{\bf{\bar H}}_{sr}^H + \sigma _s^2{{\bf{\Sigma }}_{sr}}\nonumber\\
&\quad \quad \quad + {{\sigma _{{n_{sr}}}^2{{\bf{I}}}} \mathord{\left/
 {\vphantom {{\sigma _{{n_{sr}}}^2{{\bf{I}}}} {{\rm{tr}}({\bf{F}}_s^{\left( l \right)}{\bf{F}}_s^{{{\left( l \right)}^H}})}}} \right.
 \kern-\nulldelimiterspace} {{\rm{tr}}({\bf{F}}_s^{\left( l \right)}{\bf{F}}_s^{{{\left( l \right)}^H}})}}){\bf{F}}_r^{{{\left( l \right)}^H}})\nonumber.
\end{align}

Note that for any given ${\bf{\tilde F}}_s^{\left( l \right)}$, the objective function is decreasing in ${\rm{tr}}({\bf{F}}_s^{\left( l \right)}{\bf{F}}_s^{{{\left( l \right)}^H}})$. Similarly, we can verify that the objective function also decreases with respect to ${\rm{tr}}({\bf{F}}_r^{\left( l \right)}(\sigma _s^2{{{\bf{\bar H}}}_{sr}}{\bf{F}}_s^{\left( l \right)}{\bf{F}}_s^{{{\left( l \right)}^H}}{\bf{\bar H}}_{sr}^H + \sigma _s^2\alpha _1^{\left( l \right)}{{\bf{\Sigma }}_{sr}} + \sigma _{{n_{sr}}}^2{{\bf{I}}}){\bf{F}}_r^{{{\left( l \right)}^H}})$. Thus, the optimal solutions of ${{\bf{F}}_s^{\left( l \right)}}$ and ${{\bf{F}}_r^{\left( l \right)}}$ are obtained when $\alpha _1^{\left( l \right)} = {P_s}/\sigma _s^2$ and $\alpha _2^{\left( l \right)} = {P_r}$.

\section{Derivation of (29)}
Based on the SVD and the eigenvalue decomposition (EVD), we have the following expressions
\begin{align}
& {{\bf{\Sigma }}_{sr}} = {{\bf{U}}_{{\Sigma _{sr}}}}{{\bf{\Lambda }}_{{\Sigma _{sr}}}}{\bf{U}}_{{\Sigma _{sr}}}^H,\\
& {{{\bf{\hat \Sigma }}}_{rd}} = {\bf{U}}_{{\Sigma _{rd}}}^{\left( l \right)}{\bf{\Lambda }}_{{\Sigma _{rd}}}^{\left( l \right)}{\bf{U}}_{{\Sigma _{rd}}}^{{{\left( l \right)}^H}},\\
& {\bf{\tilde \Lambda }}_{{\Sigma _{sr}}}^{\left( l \right)} = \alpha _1^{\left( l \right)}{{\bf{\Lambda }}_{{\Sigma _{sr}}}} + \sigma _{{n_{sr}}}^2{{\bf{I}}},\\
& {\bf{\tilde \Lambda }}_{{\Sigma _{rd}}}^{\left( l \right)} = \alpha _2^{\left( l \right)}{\bf{\Lambda }}_{{\Sigma _{rd}}}^{\left( l \right)} + \sigma _{{n_{rd}}}^2{{\bf{I}}},\\
& {\bf{\tilde H}}_{sr}^{\left( l \right)} \buildrel \Delta \over = {\bf{\tilde \Lambda }}_{{\Sigma _{sr}}}^{{{\left( l \right)}^{ - \frac{1}{2}}}}{\bf{U}}_{{\Sigma _{sr}}}^H{{{\bf{\bar H}}}_{sr}} = {\bf{\tilde U}}_{sr}^{\left( l \right)}{\bf{\tilde \Lambda }}_{sr}^{\left( l \right)}{\bf{\tilde V}}_{sr}^{{{\left( l \right)}^H}},\\
& {\bf{\tilde H}}_{rd}^{\left( l \right)} \buildrel \Delta \over = {\bf{\tilde \Lambda }}_{{\Sigma _{rd}}}^{{{\left( l \right)}^{{\rm{ - }}\frac{1}{2}}}}{\bf{U}}_{{\Sigma _{rd}}}^{{{\left( l \right)}^H}}{\bf{\bar H}}_{rd}^{\left( l \right)} = {\bf{\tilde U}}_{rd}^{\left( l \right)}{\bf{\tilde \Lambda }}_{rd}^{\left( l \right)}{\bf{\tilde V}}_{rd}^{{{\left( l \right)}^H}}.
\end{align}

From (76)-(79), we have
\begin{align}
&\alpha _1^{\left( l \right)}{{\bf{\Sigma }}_{sr}} + \sigma _{{n_{sr}}}^2{{\bf{I}}}{\rm{ = }}{{\bf{U}}_{{\Sigma _{sr}}}}{\rm{(}}\alpha _1^{\left( l \right)}{{\bf{\Lambda }}_{{\Sigma _{sr}}}} + \sigma _{{n_{sr}}}^2{{\bf{I}}}{\rm{)}}{\bf{U}}_{{\Sigma _{sr}}}^H\nonumber\\
&\quad \quad \quad \quad \quad \quad \quad \quad{\rm{ = }}{{\bf{U}}_{{\Sigma _{sr}}}}{\bf{\tilde \Lambda }}_{{\Sigma _{sr}}}^{\left( l \right)}{\bf{U}}_{{\Sigma _{sr}}}^H \nonumber,\\
&\alpha _2^{\left( l \right)}{{{\bf{\hat \Sigma }}}_{rd}} + \sigma _{{n_{rd}}}^2{{\bf{I}}}{\rm{ = }}{\bf{U}}_{{\Sigma _{rd}}}^{\left( l \right)}{\rm{(}}\alpha _2^{\left( l \right)}{\bf{\Lambda }}_{{\Sigma _{rd}}}^{\left( l \right)} + \sigma _{{n_{rd}}}^2{{\bf{I}}}{\rm{)}}{\bf{U}}_{{\Sigma _{rd}}}^{{{\left( l \right)}^H}}\nonumber\\
&\quad \quad \quad \quad \quad \quad \quad \quad{\rm{ = }}{\bf{U}}_{{\Sigma _{rd}}}^{\left( l \right)}{\bf{\tilde \Lambda }}_{{\Sigma _{rd}}}^{\left( l \right)}{\bf{U}}_{{\Sigma _{rd}}}^{{{\left( l \right)}^H}}\nonumber.
\end{align}
By using (80) and (81) and introducing ${\bf{\tilde F}}_r^{\left( l \right)}{\rm{ = }}{\bf{F}}_r^{\left( l \right)}{{\bf{U}}_{{\Sigma _{sr}}}}{\bf{\tilde \Lambda }}_{{\Sigma _{sr}}}^{{{\left( l \right)}^{\frac{1}{2}}}}$, the problem (26) can be equivalently written as
\begin{align}
&\min J\left( {{\bf{F}}_s^{\left( l \right)},{\bf{F}}_r^{\left( l \right)}} \right) \nonumber\\
=&\left| {{{\left( {\sigma _s^{ - 2}{\bf{I}} + {\bf{F}}_s^{{{\left( l \right)}^H}}{\bf{\tilde H}}_{sr}^H{\bf{F}}_r^{{{\left( l \right)}^H}}{\bf{\tilde H}}_{rd}^{{{\left( l \right)}^H}}{{{\bf{\tilde B}}}^{{{\left( l \right)}^{ - 1}}}}{\bf{\tilde H}}_{rd}^{\left( l \right)}{\bf{F}}_r^{\left( l \right)}{{{\bf{\tilde H}}}_{sr}}{\bf{F}}_s^{\left( l \right)}} \right)}^{{\rm{ - }}1}}} \right| \nonumber\\
& s.t. \quad \ {\rm{tr}}\left( {\sigma _s^2{\bf{F}}_s^{\left( l \right)}{\bf{F}}_s^{{{\left( l \right)}^H}}} \right) \le {P_s} \nonumber\\
& \qquad \ \ {\rm{tr}}({\bf{F}}_r^{\left( l \right)}(\sigma _s^2{{{\bf{\bar H}}}_{sr}}{\bf{F}}_s^{\left( l \right)}{\bf{F}}_s^{{{\left( l \right)}^H}}{\bf{\bar H}}_{sr}^H + \sigma _s^2\alpha _1^{\left( l \right)}{{\bf{\Sigma }}_{sr}}\nonumber \\
& \qquad \ \quad  + \sigma _{{n_{sr}}}^2{{\bf{I}}}){\bf{F}}_r^{{{\left( l \right)}^H}}) \le {P_r},\nonumber
\end{align}
\noindent where
 ${{{\bf{\tilde B}}}^{\left( l \right)}} = {\bf{\tilde H}}_{rd}^{\left( l \right)}{\bf{\tilde F}}_r^{\left( l \right)}{\bf{\tilde F}}_r^{{{\left( l \right)}^H}}{\bf{\tilde H}}_{rd}^{{{\left( l \right)}^H}} + {{\bf{I}}}$.

Using Theorem 1 in [35], we obtain a similar structure for the optimisation problem. ${\bf{\tilde F}}_r^{\left( l \right)} = {\bf{\tilde V}}_{rd}^{\left( l \right)}{\bf{\Lambda }}_r^{\left( l \right)}{\bf{\tilde U}}_{sr}^{{{\left( l \right)}^H}}$, ${\bf{F}}_s^{\left( l \right)} = {\bf{\tilde V}}_{sr}^{\left( l \right)}{\bf{\Lambda }}_s^{\left( l \right)}{\bf{V}}_o^{\left( l \right)}$, where ${\bf{V}}_o^{\left( l \right)}$  is a unitary matrix yet to be determined.
The solution to the reformulated optimisation problem is given by the source and relay precoders.
By substituting the expressions of the precoders into (23), we obtain
\begin{equation}
{{\bf{E}}^{\left( l \right)}} = {{\bf{U}}^{\left( l \right)}}{\bf{V}}_o^{{{\left( l \right)}^H}}{{\bf{\Sigma }}^{{{\left( l \right)}^{ - 1/2}}}}{{\bf{\Sigma }}^{{{\left( l \right)}^{ - 1/2}}}}{\bf{V}}_o^{\left( l \right)}{{\bf{U}}^{{{\left( l \right)}^H}}}\nonumber,
\end{equation}
\noindent where we have ${{\bf{\Sigma }}^{\left( l \right)}} \buildrel \Delta \over = (\sigma _s^{ - 2}{{\bf{I}}} + {\bf{\tilde \Lambda }}_{sr}^{{{\left( l \right)}^2}}{\bf{\Lambda }}_s^{{{\left( l \right)}^2}}{\bf{\Lambda }}_r^{{{\left( l \right)}^2}}{\bf{\tilde \Lambda }}_{rd}^{{{\left( l \right)}^2}}({\bf{\tilde \Lambda }}_{rd}^{{{\left( l \right)}^2}}{\bf{\Lambda }}_r^{{{\left( l \right)}^2}}+{{\bf{I}}}{)^{{\rm{ - }}1}}) $. We note that the lower bound of MSE is achieved when the objective function is a diagonal matrix with equal diagonal elements, namely, ${{\bf{E}}^{\left( l \right)}} = \gamma {{\bf{I}}}$.
  Then we define ${{{\bf{\tilde U}}}^{\left( l \right)}}{\rm{ = }}\bar \sigma {{\bf{U}}^{{{\left( l \right)}^{{\rm{ - }}H}}}}$ and apply the GMD to ${{\bf{\Sigma }}^{{{\left( l \right)}^{ - 1/2}}}}$ to make the diagonal entries of an upper triangular matrix all equal.
We obtain ${{\bf{\Sigma }}^{{{\left( l \right)}^{ - 1/2}}}} = {{\bf{Q}}^{\left( l \right)}}{{{\bf{\tilde U}}}^{\left( l \right)}}{\bf{\Phi }}_s^{{{\left( l \right)}^H}}$,  where ${{\bf{Q}}^{\left( l \right)}}$ and ${\bf{\Phi }}_s^{\left( l \right)}$ are unitary matrices, and ${{{\bf{\tilde U}}}^{\left( l \right)}}$ is an upper triangular matrix with equal diagonal elements $\bar \sigma$. Let ${\bf{V}}_o^{\left( l \right)} = {\bf{\Phi }}_s^{\left( l \right)}$, we obtain ${{\bf{E}}^{\left( l \right)}} = {\bar \sigma ^2}{\bf{I}}$.
 From the equation above, it can be verified that the equality is achieved.

\section{Derivation of the solution in (39) and (40)}
The Lagrangian function with respect to (37) can be written as
\begin{align}
&L = \sum\limits_{i = 1}^{{N_d}} {\ln \left( {\frac{{{y_i}\tilde \lambda _{2,i}^2{x_i}\tilde \lambda _{1,i}^2 + {y_i}\tilde \lambda _{2,i}^2 + {x_i}\tilde \lambda _{1,i}^2 + 1}}{{{y_i}\tilde \lambda _{2,i}^2 + {x_i}\tilde \lambda _{1,i}^2 + 1}}} \right)} \nonumber\\
& \quad \quad {\rm{ + }}{{\tilde \mu }_s}\left( {\sum\limits_{i = 1}^{{N_d}} {{x_i}} {\rm{ - }}{P_s}} \right){\rm{ + }}{{\tilde \mu }_r}\left( {\sum\limits_{i = 1}^{{N_d}} {{y_i}} {\rm{ - }}{P_r}} \right) \nonumber\\
& \quad \quad {\rm{ - }}\sum\limits_{i = 1}^{{N_d}} {{\upsilon _{s{\rm{,}}i}}{x_i}} {\rm{ - }}\sum\limits_{i = 1}^{{N_d}} {{\upsilon _{r{\rm{,}}i}}{y_i}} \nonumber.
\end{align}

As mentioned, if ${{x_i}}$ is given, (37) is a convex optimization problem (for ${{y_i}}$). Thus, we can obtain the optimum ${{y_i}}$ using the KKT conditions\cite{boyd2004convex}. The KKT optimality conditions for solving ${{y_i}}, {\rm{1}} \le i \le {N_s}$ are given as follows:
\begin{align}
&\frac{{\partial L}}{{\partial {y_i}}}{\rm{ =   - }}\frac{{\frac{{\tilde \lambda _{2,i}^2{x_i}\tilde \lambda _{1,i}^2\left( {{x_i}\tilde \lambda _{1,i}^2 + 1} \right)}}{{{{\left( {{y_i}\tilde \lambda _{2,i}^2 + {x_i}\tilde \lambda _{1,i}^2 + 1} \right)}^2}}}}}{{\frac{{{y_i}\tilde \lambda _{2,i}^2{x_i}\tilde \lambda _{1,i}^2 + {y_i}\tilde \lambda _{2,i}^2 + {x_i}\tilde \lambda _{1,i}^2 + 1}}{{{y_i}\tilde \lambda _{2,i}^2 + {x_i}\tilde \lambda _{1,i}^2 + 1}}}}{\rm{ + }}{{\tilde \mu }_r}{\rm{ - }}{\upsilon _{r{\rm{,}}i}}{\rm{ = }}0,\\
&{\upsilon _{r{\rm{,}}i}}{y_i}{\rm{ = }}0,\\
&{{\tilde \mu }_r}\left( {\sum\limits_{i = 1}^{{N_d}} {{y_i}} {\rm{ - }}{P_r}} \right){\rm{ = }}0,\\
&{{\tilde \mu }_r}{\rm{,\ }}{\upsilon _{r{\rm{,}}i}}{\rm{,\ }}{y_i} \ge 0.
\end{align}

Substituting (82) into (83) and considering that ${y_i} \ge 0$, we have ${\upsilon _{r{\rm{,}}i}}{\rm{ = 0}}$. After some straightforward manipulations and the use of (85), we can have the optimum ${y_i} \ge 0$ given as (39), where ${\mu _s}{\rm{ = }}{1 \mathord{\left/
 {\vphantom {1 {{{\tilde \mu }_s}}}} \right.\kern-\nulldelimiterspace} {{{\tilde \mu }_s}}}$ is chosen to satisfy the power constraint in (26).

\bibliographystyle{IEEETran}
\bibliography{references}

\begin{thebibliography}{10}
\providecommand{\url}[1]{#1}
\csname url@samestyle\endcsname
\providecommand{\newblock}{\relax}
\providecommand{\bibinfo}[2]{#2}
\providecommand{\BIBentrySTDinterwordspacing}{\spaceskip=0pt\relax}
\providecommand{\BIBentryALTinterwordstretchfactor}{4}
\providecommand{\BIBentryALTinterwordspacing}{\spaceskip=\fontdimen2\font plus
\BIBentryALTinterwordstretchfactor\fontdimen3\font minus
  \fontdimen4\font\relax}
\providecommand{\BIBforeignlanguage}[2]{{%
\expandafter\ifx\csname l@#1\endcsname\relax
\typeout{** WARNING: IEEEtran.bst: No hyphenation pattern has been}%
\typeout{** loaded for the language `#1'. Using the pattern for}%
\typeout{** the default language instead.}%
\else
\language=\csname l@#1\endcsname
\fi
#2}}
\providecommand{\BIBdecl}{\relax}
\BIBdecl

\bibitem{MMIMO}
R.~C. de~Lamare, ``{Massive MIMO Systems: Signal Processing Challenges and
  Future Trends},'' \emph{URSI Radio Science Bulletin}, Dec. 2013.

\bibitem{AFDF}
J.~Laneman, D.~Tse, and G.~Wornell, ``{Cooperative diversity in wireless
  networks: Efficient protocols and outage behavior},'' \emph{IEEE Trans. Inf.
  Theory}, vol.~50, no.~12, pp. 3062--3080, Dec. 2004.

\bibitem{relay}
X.~Tang and Y.~Hua, ``{Optimal design of non-regenerative MIMO wireless
  relays},'' \emph{IEEE Trans. Wireless Commun.}, vol.~6, no.~4, pp.
  1398--1407, Apr. 2007.

\bibitem{relayreceiver}
W.~Guan and H.~Luo, ``{Joint MMSE transceiver design in non-regenerative MIMO
  relay systems},'' \emph{IEEE Commun. Lett.}, vol.~12, no.~7, pp. 517--519,
  Jul. 2008.

\bibitem{joint}
F.~Tseng and W.~Wu, ``{Linear MMSE transceiver design in amplify-and-forward
  MIMO relay systems},'' \emph{IEEE Trans. Veh. Technol.}, vol.~59, no.~2, pp.
  754--765, Feb. 2010.

\bibitem{xing2010robust}
C.~Xing, S.~Ma, and Y.~Wu, ``{Robust joint design of linear relay precoder and
  destination equalizer for dual-hop amplify-and-forward MIMO relay systems},''
  \emph{IEEE Trans. Signal Process.}, vol.~58, no.~4, pp. 2273--2283, Apr.
  2010.

\bibitem{jointrobust}
Y.~Rong, ``{Robust design for linear non-regenerative MIMO relays with
  imperfect channel state information},'' \emph{IEEE Trans. Signal Process.},
  vol.~59, no.~5, pp. 2455--2460, May 2011.

\bibitem{erez2005capacity}
U.~Erez, S.~Shamai, and R.~Zamir, ``{Capacity and lattice strategies for
  canceling known interference},'' \emph{IEEE Trans. Inf. Theory}, vol.~51,
  no.~11, pp. 3820--3833, Nov. 2005.

\bibitem{costa1983writing}
M.~Costa, ``Writing on dirty paper (corresp.),'' \emph{IEEE Trans. Inf.
  Theory}, vol.~29, no.~3, pp. 439--441, May 1983.

\bibitem{THPnodirect}
A.~Millar, S.~Weiss, and R.~Stewart, ``{Tomlinson Harashima precoding design
  for non-regenerative MIMO relay networks},'' in \emph{Proc. IEEE Veh.
  Technol. Conf}, May 2011, pp. 1--5.

\bibitem{THPdirect}
F.~Tseng, M.~Chang, and W.~Wu, ``{Joint Tomlinson--Harashima source and linear
  relay precoder design in amplify-and-forward MIMO relay systems via MMSE
  criterion},'' \emph{IEEE Trans. Veh. Technol.}, vol.~60, no.~4, pp.
  1687--1698, May 2011.

\bibitem{Lprec}
U.~W. N. J.~A. Joham, M., ``{Linear transmit processing in MIMO communications
  systems},'' \emph{IEEE Trans. Signal Process.}, vol.~53, no.~8, pp.
  2700--2712, Aug. 2005.

\bibitem{switch}
Y.~Cai, R.~de~Lamare, and R.~Fa, ``Switched interleaving techniques with
  limited feedback for interference mitigation in ds-cdma systems,'' \emph{IEEE
  Transactions on Communications}, vol.~59, no.~7, pp. 1946--1956, July 2011.

\bibitem{switchmc}
Y.~Cai, R.~de~Lamare, and D.~Le~Ruyet, ``Transmit processing techniques based
  on switched interleaving and limited feedback for interference mitigation in
  multiantenna mc-cdma systems,'' \emph{IEEE Transactions on Vehicular
  Technology}, vol.~60, no.~4, pp. 1559--1570, May 2011.

\bibitem{Kek01}
K.~Zu and R.~C. de~Lamare, ``{Low-Complexity Lattice Reduction-Aided
  Regularized Block Diagonalization for MU-MIMO Systems},'' \emph{IEEE
  Communications Letters}, vol.~16, no.~6, pp. 925--928, June 2012.

\bibitem{Kek02}
K.~Zu, R.~C. de~Lamare, and M.~Haardt, ``{Generalized design of low-complexity
  block diagonalization type precoding algorithms for multiuser MIMO
  systems},'' \emph{IEEE Transactions on Communications}, vol.~61, no.~10, pp.
  4232--4242, October 2013.

\bibitem{jimenez2012iterative}
I.~Jimenez, M.~Barrenechea, M.~Mendicute, and E.~Arruti, ``{Iterative joint
  MMSE design of relaying MIMO downlink schemes with non-linearly precoded
  transmission},'' in \emph{Proc. IEEE SPAWC 2012}, Jun. 2012, pp. 424--428.

\bibitem{jimenez2011multiuser}
I.~Jimenez, S.~Weiss, M.~Mendicute, and E.~Arruti, ``{Multiuser MIMO
  amplify-and-forward relaying schemes with vector precoding},'' in \emph{Proc.
  IEEE ISSPIT 2011}, Dec. 2011, pp. 514--519.

\bibitem{barrenechea2012low}
M.~Barrenechea, A.~Burg, and M.~Mendicute, ``{Low-complexity vector precoding
  for multi-user systems},'' in \emph{Proc. IEEE ASILOMAR 2012}, Nov. 2012, pp.
  453--457.

\bibitem{millar2012robust}
A.~Millar, S.~Weiss, and R.~Stewart, ``{Robust transceiver design for MIMO
  relay systems with Tomlinson Harashima Precoding},'' in \emph{Proc. IEEE
  EUSIPCO 2012}, Aug. 2012, pp. 1374--1378.

\bibitem{xing2012robust}
C.~Xing, S.~Ma, F.~Gao, and Y.~Wu, ``{Robust transceiver with
  Tomlinson-Harashima precoding for amplify-and-forward MIMO relaying
  systems},'' \emph{IEEE J. Sel. Areas Commun.}, vol.~30, no.~8, pp.
  1370--1382, Sep. 2012.

\bibitem{tseng2012robust}
F.~Tseng, M.~Chang, and W.~Wu, ``{Robust Tomlinson-Harashima source and linear
  relay precoders design in amplify-and-forward MIMO relay systems},''
  \emph{IEEE Trans. Commun.}, vol.~60, no.~4, pp. 1124--1137, Apr. 2012.

\bibitem{de2008multibranch}
R.~de~Lamare and R.~Sampaio-Neto, ``{Minimum mean-squared error iterative
  successive parallel arbitrated decision feedback detectors for DS-CDMA
  systems},'' \emph{IEEE Trans. Commun.}, vol.~56, no.~5, pp. 778--789, May
  2008.

\bibitem{stmb}
Y.~Cai and R.~de~Lamare, ``Space-time adaptive mmse multiuser decision feedback
  detectors with multiple-feedback interference cancellation for cdma
  systems,'' \emph{IEEE Transactions on Vehicular Technology}, vol.~58, no.~8,
  pp. 4129--4140, Oct 2009.

\bibitem{jidf}
R.~de~Lamare and R.~Sampaio-Neto, ``Adaptive reduced-rank processing based on
  joint and iterative interpolation, decimation, and filtering,'' \emph{IEEE
  Transactions on Signal Processing}, vol.~57, no.~7, pp. 2503--2514, July
  2009.

\bibitem{jio}
------, ``Reduced-rank adaptive filtering based on joint iterative optimization
  of adaptive filters,'' \emph{IEEE Signal Processing Letters}, vol.~14,
  no.~12, pp. 980--983, Dec 2007.

\bibitem{mfsic}
P.~Li, R.~de~Lamare, and R.~Fa, ``Multiple feedback successive interference
  cancellation detection for multiuser mimo systems,'' \emph{IEEE Transactions
  on Wireless Communications}, vol.~10, no.~8, pp. 2434--2439, August 2011.

\bibitem{mbdf}
R.~de~Lamare, ``Adaptive and iterative multi-branch mmse decision feedback
  detection algorithms for multi-antenna systems,'' \emph{IEEE Transactions on
  Wireless Communications}, vol.~12, no.~10, pp. 5294--5308, October 2013.

\bibitem{did}
P.~Li and R.~de~Lamare, ``Distributed iterative detection with reduced message
  passing for networked mimo cellular systems,'' \emph{IEEE Transactions on
  Vehicular Technology}, vol.~63, no.~6, pp. 2947--2954, July 2014.

\bibitem{MBTHP}
K.~Zu, R.~de~Lamare, and M.~Haardt, ``{Multi-branch Tomlinson-Harashima
  precoding for single-user MIMO systems},'' in \emph{Proc. IEEE WSA 2012},
  Mar. 2012, pp. 36--40.

\bibitem{multibranch2014}
------, ``{Multi-branch Tomlinson-Harashima precoding design for MU-MIMO
  systems: theory and algorithms},'' \emph{IEEE Trans. Commun.}, vol.~62,
  no.~3, pp. 939--959, Mar. 2014.

\bibitem{MBIET}
R.~Fa and R.~de~Lamare, ``{Multi-branch successive interference cancellation
  for MIMO spatial multiplexing systems: design, analysis and adaptive
  implementation},'' \emph{IET Commun.}, vol.~5, no.~4, pp. 484--494, Mar.
  2011.

\bibitem{peters2009relay}
S.~W. Peters, A.~Y. Panah, K.~T. Truong, and R.~W. Heath, ``{Relay
  architectures for 3GPP LTE-Advanced},'' \emph{EURASIP J. Wireless Commun.
  Netw.}, vol. 2009, pp. 1--14, Mar. 2009.

\bibitem{fischer2002precoding}
R.~Fischer, \emph{{Precoding and Signal Shaping for Digital
  Transmission}}.\hskip 1em plus 0.5em minus 0.4em\relax New York, USA:
  Wiley-IEEE Press, 2002.

\bibitem{windpassinger2004precoding}
C.~Windpassinger, R.~Fischer, T.~Vencel, and J.~Huber, ``Precoding in
  multiantenna and multiuser communications,'' \emph{IEEE Trans. Wireless
  Commun.}, vol.~3, no.~4, pp. 1305--1316, Jul. 2004.

\bibitem{simeone2004linear}
O.~Simeone, Y.~Bar-Ness, and U.~Spagnolini, ``{Linear and nonlinear
  preequalization/equalization for MIMO systems with long-term channel state
  information at the transmitter},'' \emph{IEEE Trans. Wireless Commun.},
  vol.~3, no.~2, pp. 373--378, Mar. 2004.

\bibitem{liu2007improved}
J.~Liu and W.~Kizymien, ``{Improved Tomlinson-Harashima precoding for the
  downlink of multi-user MIMO systems},'' \emph{Can. J. Elect. Comput. Eng.},
  vol.~32, no.~3, pp. 133--144, Summer 2007.

\bibitem{dabbagh2008multiple}
A.~D. Dabbagh and D.~J. Love, ``{Multiple antenna MMSE based downlink precoding
  with quantized feedback or channel mismatch},'' \emph{IEEE Trans. Commun.},
  vol.~56, no.~11, pp. 1859--1868, Nov. 2008.

\bibitem{matrixdistribution}
A.~Gupta and D.~Nagar, \emph{{Matrix Variate Distributions}}.\hskip 1em plus
  0.5em minus 0.4em\relax London, U.K.: Chapman \& Hall/CRC, 2000.

\bibitem{exponentialmodel}
L.~Musavian, M.~Nakhai, M.~Dohler, and A.~Aghvami, ``{Effect of channel
  uncertainty on the mutual information of MIMO fading channels},'' \emph{IEEE
  Trans. Veh. Technol.}, vol.~56, no.~5, pp. 2798--2806, Sep. 2007.

\bibitem{kronecker}
M.~Ding and S.~Blostein, ``{MIMO minimum total MSE transceiver design with
  imperfect CSI at both ends},'' \emph{IEEE Trans. Signal Process.}, vol.~57,
  no.~3, pp. 1141--1150, Mar. 2009.

\bibitem{biguesh2006training}
M.~Biguesh and A.~B. Gershman, ``{Training-based MIMO channel estimation: a
  study of estimator tradeoffs and optimal training signals},'' \emph{IEEE
  Trans. Signal Process.}, vol.~54, no.~3, pp. 884--893, Mar. 2006.

\bibitem{yoo2004mimo}
T.~Yoo, E.~Yoon, and A.~Goldsmith, ``{MIMO capacity with channel uncertainty:
  Does feedback help?}'' in \emph{Proc. IEEE GLOBECOM 2004}, vol.~1, Nov. 2004,
  pp. 96--100.

\bibitem{xing2013general}
C.~Xing, S.~Ma, Z.~Fei, Y.~Wu, and H.~V. Poor, ``{A general robust linear
  transceiver design for multi-hop amplify-and-forward MIMO relaying
  systems},'' \emph{IEEE Trans. Signal Process.}, vol.~61, no.~5, pp.
  1196--1209, Mar. 2013.

\bibitem{matrixinverse}
D.~Bernstein, \emph{{Matrix Mathematics: Theory, Facts, and Formulas}}.\hskip
  1em plus 0.5em minus 0.4em\relax Princeton, USA: Princeton Univ. Press, 2011.

\bibitem{larsson2008space}
E.~G. Larsson and P.~Stoica, \emph{Space-time block coding for wireless
  communications}.\hskip 1em plus 0.5em minus 0.4em\relax Cambridge, U.K.:
  Cambridge univ. press, 2008.

\bibitem{diagonal}
D.~Palomar, J.~Cioffi, and M.~Lagunas, ``{Joint Tx-Rx beamforming design for
  multicarrier MIMO channels: A unified framework for convex optimization},''
  \emph{IEEE Trans. Signal Process.}, vol.~51, no.~9, pp. 2381--2401, Sep.
  2003.

\bibitem{unified}
Y.~Rong, X.~Tang, and Y.~Hua, ``{A unified framework for optimizing linear
  nonregenerative multicarrier MIMO relay communication systems},'' \emph{IEEE
  Trans. Signal Process.}, vol.~57, no.~12, pp. 4837--4851, Dec. 2009.

\bibitem{boyd2004convex}
S.~Boyd and L.~Vandenberghe, \emph{{Convex Optimization}}.\hskip 1em plus 0.5em
  minus 0.4em\relax Cambridge, U.K.: Cambridge univ. press, 2004.

\bibitem{yu2004iterative}
W.~Yu, W.~Rhee, S.~Boyd, and J.~M. Cioffi, ``{Iterative water-filling for
  Gaussian vector multiple-access channels},'' \emph{IEEE Trans. Inf. Theory},
  vol.~50, no.~1, pp. 145--152, Jan. 2004.

\bibitem{gmd}
Y.~Jiang, J.~Li, and W.~Hager, ``{Joint transceiver design for MIMO
  communications using geometric mean decomposition},'' \emph{IEEE Trans.
  Signal Process.}, vol.~53, no.~10, pp. 3791--3803, Oct. 2005.

\bibitem{matrixcomputations}
G.~Golub and C.~Van~Loan, \emph{{Matrix Computations}}.\hskip 1em plus 0.5em
  minus 0.4em\relax Baltimore, USA: Johns Hopkins Univ. Press, 1996.

\bibitem{windpassinger2003precoding}
C.~Windpassinger, T.~Vencel, and R.~F. Fischer, ``{Precoding and loading for
  BLAST-like systems},'' in \emph{Proc. IEEE ICC 2003}, vol.~5, May 2003, pp.
  3061--3065.

\bibitem{wubben2002efficient}
D.~W{\"u}bben, J.~Rinas, R.~B{\"o}hnke, V.~K{\"u}hn, and K.~Kammeyer,
  ``{Efficient algorithm for detecting layered space-time codes},'' in
  \emph{Proc. ITG Conference on Source and Channel Coding (SCC)}, Jan. 2002,
  pp. 1--7.

\bibitem{habendorf2006ordering}
R.~Habendorf and G.~Fettweis, ``{On ordering optimization for MIMO systems with
  decentralized receivers},'' in \emph{Proc. IEEE VTC 2006-Spring}, vol.~4, May
  2006, pp. 1844--1848.

\end{thebibliography}

\end{document}